\documentclass{emulateapj}

\begin{document}

\title{The progenitors of the compact early-type galaxies at high-redshift}

\author{Christina C. Williams\altaffilmark{1}, 
Mauro Giavalisco\altaffilmark{1}, 
Paolo Cassata\altaffilmark{2}, 
Elena Tundo\altaffilmark{3}, 
Tommy Wiklind\altaffilmark{4}, 
Yicheng Guo\altaffilmark{5}, 
Bomee Lee\altaffilmark{1}, 
Guillermo Barro\altaffilmark{5}, 
Stijn Wuyts\altaffilmark{6}, 
Eric F. Bell\altaffilmark{7}, 
Christopher J. Conselice\altaffilmark{3}
Avishai Dekel\altaffilmark{8}, 
Sandra Faber\altaffilmark{5}, 
Henry C. Ferguson\altaffilmark{9}, 
Norman Grogin\altaffilmark{9}, 
Nimish Hathi\altaffilmark{10}, 
Kuang-Han Huang\altaffilmark{11}, 
Dale Kocevski\altaffilmark{12}, 
Anton Koekemoer\altaffilmark{9}, 
David C. Koo\altaffilmark{5},
Swara Ravindranath\altaffilmark{13}
Sara Salimbeni\altaffilmark{14}
}

\altaffiltext{1}{Department of Astronomy, University of Massachusetts, 710 North Pleasant Street, Amherst, MA 01003, USA, ccwillia@astro.umass.edu}
\altaffiltext{2}{Aix Marseille Universit\'e, CNRS, LAM (Laboratoire
  d'Astrophysique de Marseille) UMR 7326, 13388, Marseille, France} 
\altaffiltext{3}{The School of Physics and Astronomy, University of Nottingham, Nottingham, NG7 2RD UK}
\altaffiltext{4}{Joint ALMA Observatory, ESO, Santiago, Chile}
\altaffiltext{5}{UCO/Lick Observatory, Department of Astronomy and
  Astrophysics, University of California, Santa Cruz, CA 95064, USA}
\altaffiltext{6}{Max-Planck-Institut für Extraterrestrische Physik (MPE), Postfach 1312, D-85741 Garching, Germany}
\altaffiltext{7}{Department of Astronomy, University of Michigan, 500 Church Street, Ann Arbor, MI 48109, USA}
\altaffiltext{8}{Racah Institute of Physics, The Hebrew University, Jerusalem 91904, Israel}
\altaffiltext{9}{Space Telescope Science Institute, 3700 San Martin
  Boulevard, Baltimore, MD 21218, USA} 
\altaffiltext{10}{Carnegie Observatories, Pasadena, CA 91101, USA}
\altaffiltext{11}{Department of Physics and Astronomy, Johns Hopkins University, 3400 North Charles Street, Baltimore, MD 21218, USA}
\altaffiltext{12}{Department of Physics and Astronomy, University of Kentucky, Lexington, KY 40506, USA}
\altaffiltext{13}{Inter-University of Astronomy \& Astrophysics, Pune, Maharashtra, India}
\altaffiltext{14}{Marlboro College, Marlboro, VT}

\begin{abstract}

We use GOODS and CANDELS images to identify progenitors of massive
($M>10^{10}M_{\odot}$) compact ``early--type'' galaxies (ETGs) at
$z\sim1.6$. Since merging and accretion increase the size of the stellar
component of galaxies, if the progenitors are among known star-forming
galaxies, these must be compact themselves. We select candidate progenitors
among compact Lyman--break galaxies at $z\sim3$ based on their mass,
SFR and central stellar density and find that these account for a large
fraction of, and possibly all, compact ETGs at $z\sim1.6$. We find that the average
far--UV SED of the candidates is redder than that of the non-candidates, but
the optical and mid--IR SED are the same, implying that the redder UV of the
candidates is inconsistent with larger dust obscuration, and consistent with
more evolved (aging) star-formation. This is in line with other evidence that
compactness is a sensitive predictor of passivity among high--redshift massive
galaxies. We also find that the light distribution of both the compact ETGs and their candidate progenitors does not show any
extended ``halos'' surrounding the compact ``core'', both in individual images
and in stacks. We argue that this is generally inconsistent with the
morphology of merger remnants, even if gas--rich, as predicted by N--body
simulations. This suggests that the compact ETGs formed via highly dissipative, mostly
gaseous accretion of units whose stellar components are very small and
undetected in the HST images, with their stellar mass assembling in--situ, and
that they have not experienced any major merging until the epoch of
observations at $z\sim1.6$.

\end{abstract}


\section{Introduction}
\label{Introduction}

A generic prediction of the standard cosmological paradigm is that small
structures form first while big ones are assembled later by hierarchical
merging. Because the power spectrum is not truncated at any scales relevant to
galaxy formation as it evolves, early massive galaxies are comparatively much
rarer than galaxies of the same mass that assembled later. These later massive
galaxies have assembled their stellar bodies in ways and over time scales that
are rather different from the early ones and thus must generally have
different properties. Thus, the discovery of old and massive galaxies at high
redshift that have rather different structural properties than those of most
early type galaxies (ETGs) in the second half of the Hubble time is interesting
because of the possibility it offers to directly explore additional mechanisms 
of formation of massive galaxies in general, and of quenching of the star 
formation in particular.

Passive galaxies with large stellar mass, e.g. M $\gtrsim$ 10$^{10}$ M$_{\odot}$,
have been identified at redshift as high as $z\sim 3$, $\approx 16$\% of the
current age of the universe. A striking characteristic of these young members
of the population of ``ETGs'' is that they are often of very small
size, up to $\sim 5$ times smaller than galaxies of comparable mass in the
local universe, and hence have very high stellar density, up to two orders of
magnitude higher than local counterparts \citep[see, for example, ][]{Daddi2005,
  Trujillo2006, Bundy2006, Cimatti2006, vanderwel2008,
  vanDokkum2008, saracco2009, Bezanson2009, Damjanov2009, RWilliams2010,
  vanDokkum2010, Saracco2010, Cassata2011, Guo2012, Ryan2012, Cassata2013}.

In fact, at $z\sim 1.6$ the population of ETGs is
dominated by the compact ones, with more than $\gtrsim 80$\% of them having
size smaller than the lower $1\sigma$ of local ETGs of the
same mass \citep{Cassata2011,Cassata2013}. In the local universe, these
compact ETGs seem to be exceedingly rare, although there is still
ongoing debate on whether this apparent paucity is real or is due to bias in
local surveys, such as the SDSS \citep{Bournaud2007, Hopkins2009, Taylor2010,
  Mclure2012, Ragone2011, Oogi2012, Nipoti2012}

Given this apparent spectacular evolution, it is no surprise that a lot of
effort went into exploring the possible evolutionary mechanisms that have
driven it. For example, it has been suggested that individual compact
ellipticals might form extended stellar halos either by in-–situ star
formation or by dry merging and accretion \citep{Naab2007, vanDokkum2010,
  Whitaker2010, Nipoti2012, Oser2012}. In 
particular it has also been proposed that interactions and repeated minor
merging events, even if they do not increase the stellar mass by a large
amount, can energize the innermost stellar orbits and ``puff up'' the compact
galaxies \citep{Newman2012}. Concurrently, the size evolution of the
population of ETGs as a whole can also be driven by the addition
of new members coming from the late quenching of massive, large galaxies
\citep{Valentinuzzi2010a,Valentinuzzi2010b, Cassata2011, Newman2012,
Poggianti2013, Carollo2013}. In fact, from the analysis of the
evolution of ETGs of different stellar densities as a
function of redshift, \citet{Cassata2013} conclude that the addition of new, larger, ETGs is required to explain the overall increase in their numbers
from $z\sim 1$ to the present \citep{Ilbert2010, Pozzetti2010}. In
any case, it is important to realize that an accurate census of compact
galaxies in the local universe is still missing, since the SDSS samples are very likely biased against such systems \citep{Scranton2002, Valentinuzzi2010a, Cassata2013, Carollo2013} and also because the descendants of the compact galaxies might not
be easily recognized in the local universe if they became the core of systems
that developed extended stellar halos (e.g. \citet{Kormendy2009, Nipoti2012, Huang2012}).

Regardless of the problem of their subsequent evolution, however, it seems
clear that compact ETG were very abundant at high redshift, and in fact
largely dominate the population of passive galaxies at redshift $z\sim
1.2$--2.8 (Cassata et~al. 2013), at least at large mass ($M>10^{10}$
M$_{\odot}$). This raises the question of how such massive systems could form
in such a relatively short time. There are indications that whatever process
is responsible for quenching star formation in massive galaxies, it is largely
controlled by the star formation rate, with more actively star--forming
galaxies being more likely to quench \citep[e.g.][]{Peng2010, Peng2012}, and
that more compact systems are more likely to quench more effectively than
those with more diffuse mass distribution \citep{Bell2012}. But even before
they quench, the problem of how massive galaxies with such high stellar
density could form, and why they dominate the population of massive passive
galaxies at high redshift, is interesting because it seems to imply a
specific formation mechanism different from other galaxies. Is the physics
that shuts off star-formation in high-redshift galaxies producing {\it only}
compact remnants as a by-product?  Or, does it preferentially affect those
galaxies with high stellar densities?  One popular mechanism to both shut off
the star formation in a galaxy and also produce spheroidal morphologies are
major mergers \citep{Barnes1992, Hernquist1992, Hernquist1993,
  Springel2005}. Evidence of this mechanism is seen in the local universe
\citep{Sanders1988, SandersMirabel1996}, and evidence of merging has been
observed out to high redshift \citep{Lotz2008, Robaina2010, Kartaltepe2010, Kartaltepe2012}. During a merging event, however, a substantial fraction of
the pre--existing stars of the merging galaxies are dispersed to larger radii,
and its difficult to produce compact remnants unless the progenitors
themselves are also very compact, and in any case the fraction of stars
scattered to large radii is not negligible \citep{Ostriker1980, Naab2007,
  Naab2009}. Gas rich mergers may produce remnants with a very compact core
through in-situ star-formation thanks to the highly dissipative nature of the
gas \citep{KhochfarSilk2006, Hopkins2008a, Wuyts2010, Bournaud2011}, but the
pre-existing stars are still dispersed to large radii, and the gas fractions
must be high (e.g.  $\gtrsim 60$--70\%) in order to produce a large fraction
of the stellar mass in a compact remnant \citep[][]{Hopkins2008a,
  Hopkins2009b, Wuyts2010}. While it is possible that current data have not
yet probed the low surface brightness regions surrounding compact ETGs to the 
sensitivity required to to rule out evidence of major
merger activity, tidal debris, or dispersed stars, there is some general
evidence that these galaxies are truly compact in size, with no diffuse or
extended structure surrounding them \citep{vanDokkum2008, Szomoru2012,
  Bezanson2009}. There is still much debate on the role of major merging in the
buildup of the stellar mass of massive galaxies in general, regardless if
compact or not, \citep{Bell2006, Robaina2010, Genel2010, RWilliams2011, Conselice2013} and in particular
for morphologically selected spheroidal galaxies \citep{Bundy2007, Bundy2009, Kaviraj2012a, Kaviraj2012b}.

Theoretical work have shown that violent disk instability \citep[VDI;][]{
  Dekel2009b}, driven by intense accretion of cold gas from the cosmic web
\citep{BirnboimDekel2003, Keres2005, Keres2009, DekelBirnboim2006, Dekel2009},
can lead to the formation of compact massive galaxies.  The gas-rich disks are
Toomre unstable, with large-scale transient perturbations and massive bound
clumps. The mutual interactions between these perturbations exert torques
that drive angular-momentum out and mass in, in the form of clump migration
and gas inflow in the inter-clump disc medium \citep{Bournaud2011, Dekel2013,
  Forbes2013}.  As long as the inflow rate is more rapid than the
star-formation rate in the disk, the inflow to the center is gas rich, and the
product is compact (Dekel \& Burkert, in preparation). The induced central
starburst can eventually lead to quenching, by gas consumption into stars
\citep{DiamondStanic2012}, by outflows driven by stellar or AGN feedback
\citep{Springel2005}, or by morphological quenching \citep{Martig2009}.  The
star formation quenching may also be related to the shutting off of cold gas
supply. Theory, confirmed by simulations, predict that after $z \sim 2$, for
dark matter halos with masses of $\sim 10^{12} M_\odot$ and above, the
incoming gas is heated by a virial shock that can be supported because of the
long cooling times \citep{DekelBirnboim2006}.

The extent to which these processes affect the formation of compact ETGs at $z>2$
remains unconstrained. Thus, progress is likely to come from the
identification and empirical studies of their progenitors before and while
they quench, namely  while they are still in the star--forming
phase and when they shut it down. Some have proposed potential progenitors
based on matching the stellar mass and the volume density of the ETGs with
those of star--forming galaxies together with assumptions of possible
evolutionary paths \citep{Whitaker2012, Barro2013, Patel2013, Stefanon2013}. \citet{Stefanon2013} in particular have identified progenitors of the most massive compact ETGs (M$>10^{11}$ M$_{\odot}$) among the most massive (M$>10^{10.6}$ M$_{\odot}$) galaxies at z$>$3 by projecting their observed stellar masses and SFRs assuming various SFHs.
But whether or not the morphological properties, star--formation rate and
stellar mass of the  more general population of putative progenitors were consistent with the compactness
of their descendants  among all passive galaxies at z$\sim$2 and their specific star formation rate have not been
considered in detail, something we set to do here.

In this paper, we try to answer the following question. Since we do not know
of any physical mechanism capable to transform a non compact stellar system
into a compact one, do star--forming galaxies exist at suitably high redshifts
(i.e. such that there is time for quenching) that are {\it plausible}
progenitors of the compact and massive ETG, namely that are compact themselves
and with stellar mass and star--formation rate such to plausibly explain their
descendants at $\langle z\rangle\sim 1.6$?  To answer this question, we
identify star-forming galaxies at $z\sim 3$ that, if they quench, can
reproduce both the mass and the stellar density, and hence size, of the
observed compact galaxies at $\langle z\rangle\sim 1.6$. In other words, we
try to see if we can identify the progenitors based on the {\it evolutionary
  consistency} assuming only that 1) the star--forming galaxies quench early
enough to be passive at $z\sim 1.6$ and 2) that no (unknown) physical
mechanism transform non-compact stellar systems into compact ones. With a
sample of plausible progenitors, we then can compare their properties to those
of other star--forming galaxies that are not plausible progenitors and see if
there are differences that might offer some insight into the formation of the
ETGs.  We present the sample, and its selection, in sections 2 and 3. In
section 4, we study the properties of these plausible progenitors, and compare
them with the rest of the normal star-forming galaxy population at $z > 3$, and we investigate
the distribution of stellar populations of different ages in the galaxies.
In section 5 we discuss the implications of our results for the evolution of
compact ETGs, in the context of the evolutionary drivers and quenching
mechanisms affecting these galaxies. Throughout this paper we assume a
cosmology with $\Omega_{\Lambda}=0.7$, $\Omega_{m}=0.3$, and $H_{o}=70$km
s$^{-1}$ Mpc$^{-1}$.

\section{Data}
\label{Data}
\subsection{Multi-wavelength Imaging and Photometry}

In this paper we use data from the Great Observatories Origins Deep Survey
\citep[GOODS;][]{Giavalisco2004}, and 4-epoch depth observations with {\it HST}/WFC3
from the Cosmic Assembly Near-infrared Deep Extragalactic Legacy Survey
(CANDELS). This covers 113 square arc-minutes of the GOODS-South field, which
includes the CANDELS Deep region  \citep{Grogin2011, Koekemoer2011} and the early release science (ERS) \citep{Windhorst2011}. The H-band images in the 4-epoch deep and ERS regions reach 1-$\sigma$ fluctuations of 26.6 and 26.3 AB arcsec$^{-2}$, respectively.

In total we make use of panchromatic photometry in GOODS-South, including
U--band data from the Visible Multiobject Spectrograph (VIMOS) on the Very
Large Telescope \citep[VLT;][]{Nonino2009}, HST/ACS B,V,i,z-band, HST/WFC3
J, H-band, VLT/ISAAC K$_{s}$ photometry \citep{Retzlaff2010}, and {\it
  Spitzer}/IRAC 3.6, 4.5, and 5.7 $\mu$m imaging (M. Dickinson et al., in
preparation), {\it Spitzer}/MIPS 24$\mu$m imaging (M. Dickinson et al., in preparation), and
GOODS-{\it Herschel}/PACS 100$\mu$m imaging \citep{Elbaz2011}.

We measure photometry (in IRAC channels and blue-ward) for galaxies in the
4-epoch CANDELS data using the object template-fitting method
\citep[TFIT;][]{Laidler2007} software package, which allows us to construct
spectral energy distributions (SEDs) with mixed-resolution datasets. All
details about the construction of the multi-wavelength photometry constructed
using TFIT is discussed in \citet{Guo2013}.

\subsection{LBG sample selection}

\begin{figure} [!t]
\begin{center}
\includegraphics[scale=0.35]{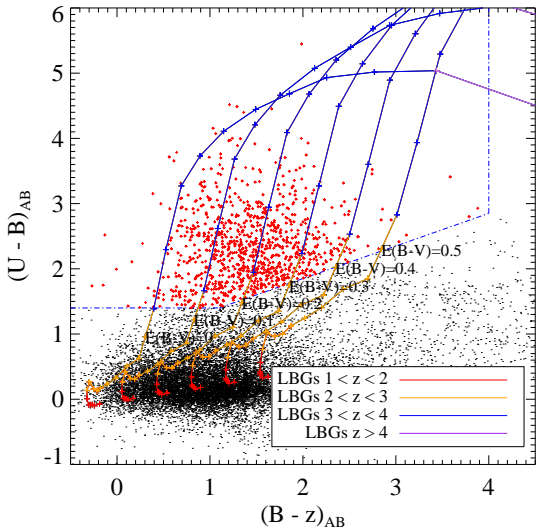}
\caption{LBG color selection. Generic LBG SED tracks (see text) plotted with varying E(B-V).}
\label{LBGsel}
\end{center}
\end{figure}

\begin{figure} [!t]
\begin{center}
\includegraphics[scale=0.51]{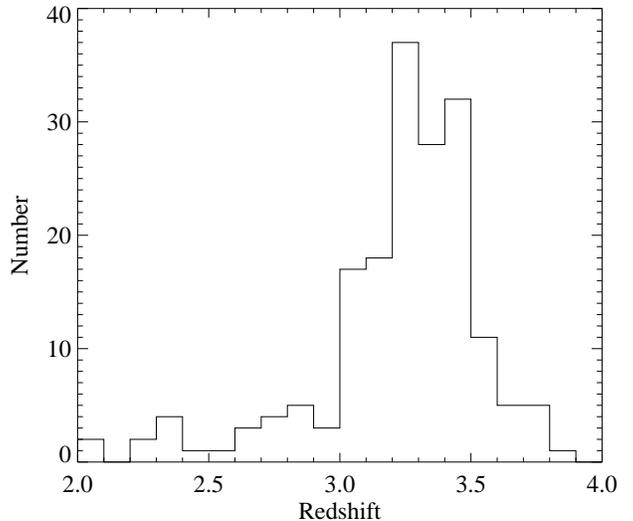}
\caption{Redshift distribution for H $<$ 25 LBGs.}
\label{zdist}
\end{center}
\end{figure}

The colors and ages of the high-redshift ETGs are such that they should be star-forming at $ z \sim 3 $ \citep{Daddi2005, Kaviraj2012a, Onodera2012}.
Therefore we select star-forming galaxies at redshift $ z \sim 3 $ from the ACS z-band
imaging using the Lyman-break color selection \citep{SteidelHamilton1993,
  Giavalisco2002}, including z-band detections with z $\gtrsim$ 26.5 (AB magnitudes). The ACS z-band is $\sim$ 90\% complete down to 26.5 for galaxies with size less than 0.3 arcseconds in half-light radius \citep{KHHuang2013}. Our U--band dropout selection criteria are

\begin{eqnarray}
(U_{VIMOS} - B_{435}) &\geq& 0.85 + 0.5 \times (B_{435}-z_{850}) \wedge  \nonumber \\
(U_{VIMOS} - B_{435}) &\geq& 1.4 \wedge B_{435}-z_{850} \leq 4,  \nonumber
\end{eqnarray}

where $\wedge$ refers to the logical ''and'' operator. We additionally require
signal to noise of at least 3 in the $B_{435}$ and $z_{850}$ bands to ensure
robust color measurements. We calibrate our U--band dropout selection using
redshifted, continuously star-forming stellar population synthesis models of
\citet{BruzualCharlot2003} with varying values of dust extinction, following
\citet{Calzetti2000}. The LBG selection, with these varying stellar population
models, are shown in Figure \ref{LBGsel}, and includes 943 objects. We
additionally remove from the LBG sample those galaxies without unique
WFC3 H-band detections (36\% of the original sample), and those with
photometric redshifts less than 2 (an additional 14\%), to ensure only
the most robust sample of z$>$3 galaxies are used in the following
analysis. Our final sample of LBGs includes 517 galaxies, 180 
are H $<$ 25, and all of which have z$\le 26.5$. As we shall see later, a
sample of candidate progenitors of compact massive early--type galaxies will
be selected among the most compact LBG which, given the sensitivity of the
GOODS images, is $\gtrsim 85$-90\% complete down to z$\le 26.5$. The
  redshift distribution for H $<$ 25 LBGs is shown in Figure \ref{zdist} .

\subsection{Measuring physical properties of LBGs}

We measure photometric redshifts and stellar masses for our sample of LBGs by
fitting stellar population synthesis models to their observed
SEDs. Photometric redshifts are derived using PEGASE 2.0 \citep{Fioc1997},
where we integrate the probability distribution function of redshift to derive
the photometric redshift.  20$\%$ of our LBGs have spectroscopically confirmed
redshifts \citep[][Stern et al. in prep]{Cristiani2000, LeFevre2004,
  Szokoly2004, Vanzella2005, Mignoli2005, Ravikumar2007, Vanzella2008,
  Kurk2009, Popesso2009}.  Using these photometric redshifts (or spectroscopic
where available), stellar masses are derived through fitting the 
stellar population synthesis models of \citet{Bruzual2007} with a
Salpeter initial mass function, as described in \citet{Guo2012}. The models
also use the Calzetti dust extinction law \citep{Calzetti2000} and the Madau
(1995) cosmic opacity, and a number of star--formation histories including
exponentially decreasing ($\tau$-models with varying time scale $\tau$),
constant, and two-component (delay) models comprised of linearly increasing
and exponentially decreasing components \citep[e.g.][]{JLee2010}. While we
found that there generally is good quantitative agreement between the stellar
mass derived using these three star formation histories, we ended up adopting
the exponentially declining or constant star--formation history that minimizes the $\chi^{2}$.

We measure SFRs of our LBGs using the observed slope of the rest--frame
ultraviolet (UV) SED, i.e. we do not use the SFR derived from the SED fitting
procedure. We make use of the correlation between the dust obscuration and the
slope of the rest--frame UV SED of starburst galaxies \citep{Meurer1999} to
derive the dust-corrected UV luminosity, and subsequently the dust-corrected UV SFR using the conversion factor by \citet{Madau1998}.

Morphological measurements in the H-band (restframe-optical at $z\sim3$) of
our LBGs are performed using the GALFIT package \citep{Peng2002}. Nearby
objects are masked using segmentation maps produced by SExtractor in the same
configuration as that used for the initial source detection. We fit light
profiles using a Sersic model. We use a PSF constructed from an average of
unsaturated stars. To ensure robust morphological measurements, we remove
galaxies from our sample if GALFIT indicated the morphological measurement was
questionable, namely such that we could not confidently rule out that they
were stars, or the signal to noise in the H-band photometry was less than
15. Previous investigations of GALFIT measurements with low signal to noise
data indicate that a signal to noise of at least 10 is required to produce
unbiased measurements \citep{Ravindranath2006, Trujillo2007,
  Cimatti2008}. These criteria remove $5\%$ of our LBG sample, 2 LBGs on the
basis of GALFIT error, and 8 LBGs on the basis of low signal-to-noise in the
H-band.

Our study uses a measure of circularized half-light radius,
defined as $R_{e} = r_{e}\sqrt{b/a}$ where $r_{e}$ is the length of the
semi-major axis in arc-seconds, and $b/a$ is the axis ratio of the galaxy. All
measurements of circularized half-light radius are converted to kpc using the
photometric redshift of the object (or spectroscopic redshift if
available). The uncertainty on the physical size of each galaxy is taken to be
20$\%$ of the measurement, based on the simulations by \citet{Cassata2011}, as
the error returned by GALFIT does not include systematic errors.

\section{Identification of plausible progenitors of compact ETGs}

Our goal in this section is to see if among star--forming galaxies at redshift
$z\sim3$, e.g. U--band dropout LBGs, there are some that are plausible
progenitors of the compact ETGs at $\langle z\rangle\sim 1.6$. 

The redshift range between $z\sim 3$ and 1.6 encompasses about 2 Gyr, so
essentially {\it every} LBG at $z\sim 3$ that quenches its star formation soon
and quickly enough after the epoch of the observation will satisfy our
definition of ``passive galaxy''.  But while quenching is an obvious
necessary condition to be classified as passive, it is not sufficient, as not
every quenched star-forming galaxy will have 1) the stellar mass, 2) the
specific star--formation rate, and 3) the stellar density of the compact,
massive and passive galaxies that we are considering here, i.e. those selected
using the criteria by \citet{Cassata2013} Thus, to identify the likely
progenitors of this specific group of galaxies, let's first exactly define
what we mean by {\it massive, passive, compact galaxies}.

We use the same criteria adopted by Cassata et~al. (2011, 2013) to define
their sample. Specifically, galaxies were selected to be in the redshift range
$1.2 < z < 2.8$ by means of photometric redshift (or spectroscopic ones when
available), and for having stellar mass $M^{*}>10^{10}$ $M_{\odot}$, and
specific star--formation rate $log_{10}(sSFR)<-2$ Gyr$^{-1}$. In the case of the ETG sample, both stellar
mass and $sSFR$ have been estimated from SED fitting to stellar population
synthesis models, assuming the star formation history as an exponentially
declining one, which is appropriate for the case of galaxies that are
completing the cessation of their star formation activity. We note that our
requirement of passivity is a relatively stringent one, with the $sSFR$ limit
being about 1/10 of the current value of the Milky Way. Recent selections of
passive galaxies in the literature use sSFR thresholds which are an order of
magnitude or even higher than our criteria \citep[e.g.][]{Barro2013}. This is
required, in our opinion, to clearly distinguish a passive galaxy, like
present--day ellipticals, from relatively low--level but continuous star
formation, such as that of massive disks, since the star--formation history of
these two types of galaxies are very different.

Compact galaxies are defined in terms of their size and stellar mass, hence
average stellar--mass surface density within the effective (half-light) radius
$r_{e}$, namely $\Sigma_{50}=\frac{0.5M_{*}}{\pi r_{e}^{2}}$. To be classified
as compact, passive galaxies of given mass must have stellar--mass surface
density larger than the value that exceeds the $1-\sigma$ scatter of the
stellar mass - size relationship for local ($z=0$) ETGs of the same mass
(thus, by definition $\approx 17$\% of local ETG are compact).  Ultra--compact
galaxies are those that are 0.4 dex smaller than the $1-\sigma$ value. In
terms of projected density, the local mass--size relationship is roughly
parallel to the line of constant $\Sigma_{50}$, so that the above definitions
translate into the following conditions: $\Sigma_{50}\gtrsim 3\times 10^{9}$
M$_{\odot}$ kpc$^{-2}$ for compact galaxies and $\Sigma_{50}\gtrsim 1.2\times
10^{10}$ M$_{\odot}$ kpc$^{-2}$ for ultra--compact ones (see Cassata
et~al. 2011). We adopt these more general classifications in the remainder of this study.

The sample by \citet{Cassata2013} includes a total of 107 ETGs with stellar
mass $M>10^{10}$ M$_{\odot}$ in the range $1.2<z<2.8$, with average redshift
$\bar z=1.6$, of which 76 are compact according to the above definition, 42 of
which are also ultra--compact, and the remaining 31 are normal ETGs,
i.e. within the $1-\sigma$ scatter in the mass--size relationship observed at
$z=0$. At high redshift ultra--compact galaxies appear to dominate the
population of massive passive galaxies. Of the 21 galaxies of this sample that
have $z\ge 2$, 4 are normal, 3 are compact and 14 are ultra--compact, suggesting
that compact and ultra--compact galaxies dominate the population of passive
galaxies at high redshift and thus were the first to become passive.

\subsection{Consistent Star-Formation Histories}

As the mass-size distributions of galaxies in the top panel of Figure
\ref{sizevsm} show, the stellar masses of the LBGs are significantly smaller
than that of the passive ones. Can then the progenitors of the compact
ETGs be found among them? The LBGs are star--forming galaxies and thus continue
to increase their stellar mass until they quench. Thus, the question can be
reformulated as whether there is enough time between $z\sim 3$ and $z\sim 1.6$
for the LBGs to cessate their star--formation activity, reach a $sSFR$ low
enough to satisfy the definition of passivity given above, and build up enough
stellar mass to reproduce the distribution of the ETGs. In the Appendix we
discuss possible quenching paths that the LBGs need to follow to be classified
as passive at $z\sim 1.6$ according to our definition. To summarize, from the
knowledge of the stellar mass and the star--formation rate of the individual
galaxies at the time of the observation, and assuming a functional form for
the star-formation history during the quenching phase, we can predict the
final stellar mass and $sSFR$ of the former LBG once they have quenched.
 This calculation shows that there are indeed plausible quenching
scenarios that could evolve some of the LBGs in our sample into passive
galaxies as defined above. Since the quenching phase is believed to be fairly
rapid \citep[e.g.][]{Peng2012}, in the calculation we model it with a
decreasing exponential function $exp(-t/\tau)$,  with the time--scale $\tau$ equal to 100 Myr for all galaxies. This timescale was estimated using the sound--crossing time in compact galaxies (we used $v_s=20$ km~s$^{-1}$ for the sound speed in the ISM of LBGs, as discussed in more detail in the Appendix).
 We also assumed that
the declined phase of the SFR began at the time of observation,
i.e. $t_q=t_{obs}$.

As discussed in the Appendix, the time difference between $z\sim 3$ and $z\sim
1.6$ is sufficiently large that from the point of view of the selection of the
candidate progenitors, these are conservative assumptions since both a later
quenching time $t_q$ and a longer $\tau$ would still result in galaxies that
are passive according to our criterion while yielding larger stellar mass and
hence increases the number of candidates.

\subsection{Progenitor Morphologies}

But just comparing the two stellar mass distributions of descendants and
candidate progenitors (after they have finished forming stars) is not
sufficient to set up physically motivated selection criteria. A key property
to consider when trying to identify the progenitors of the compact ETGs is the
morphology of these systems, and specifically their very high stellar
density. The question we are posing is: can star--forming galaxies at $z\sim
3$ with any morphology be the progenitors of the compact passive ones? Or only
galaxies with certain morphology can evolve into such systems, either via
merging or via in--situ star formation?

Based on N--body simulations of binary merging events, \citet{Wuyts2010}
suggested that the compact ETG progenitors are to be searched for among
star--forming galaxies that are compact themselves and also have large gas
fractions, i.e. $f_{gas}\gtrsim 60$\%. These progenitors in turn can form most
of the stellar mass of the final descendant in a highly concentrated
merger--triggered burst. \citet{Barro2013} also argue that a powerful
nuclear starburst in a merger remnant producing most of the stellar mass of the
remnant itself will result in a galaxy that, once passive, will resemble the
compact ones. As we will discuss later, however, it is reasonable to believe
that, whether most of the final stellar mass is formed during a
merger--induced nuclear burst or in--situ star formation, the progenitors must
be compact themselves, and must have stellar densities comparable to or higher
than the descendants, and with similar stellar mass profiles. In other
words, it is not reasonable to expect that star--forming galaxies at $z>2$
whose light profile is more diffuse than the $z\sim 1.6$ compact passive ones
can be their progenitors, even if most of the stellar mass of the final
descendant is produced after the epoch of observation in a compact
region. This is true both for the case of a single galaxy that forms stars and
quenches star formation, in situ, or for galaxies that merge.

The physical reason is that an existing diffuse stellar component cannot be
shrunk into a compact one and it cannot be hidden from observations either,
regardless if a new, more massive compact stellar component is subsequently
created (note that mergers puff up compact structures into diffuse ones, not
vice versa). To first order, if a diffuse component is observed in the CANDELS
H--band images in a star--forming galaxy at $z\sim 3$, the same diffuse
component will also be observed in the same image if the same galaxy were
placed at $z\sim 1.6$ after quenching its star formation activity. This is in
part because of the lower redshift and in part because of the increase in
stellar mass between the time when the galaxy is observed and when the
quenching process is complete.

\subsection{Progenitor Selection}

Given that the quenching phase is not instantaneous, candidate progenitors 
must have 1) smaller stellar mass than the passive ones, but
star--formation rate such that after they quench the final stellar mass
density reproduces that of the passive ones; 2) morphology and size
similar to that of the ETGs, since the ongoing formation of stars
does not change the dynamics of the pre--existing stellar orbits and hence the
appearance of the galaxies. Here we will explore a scenario where these progenitors, therefore, essentially evolve
at constant size by converting gas into stars while their stellar mass and
density increases, therefore maintaining the compact light profile observed in
the compact ETG.

\begin{figure}[!t]
\begin{center}
\includegraphics[scale=0.35]{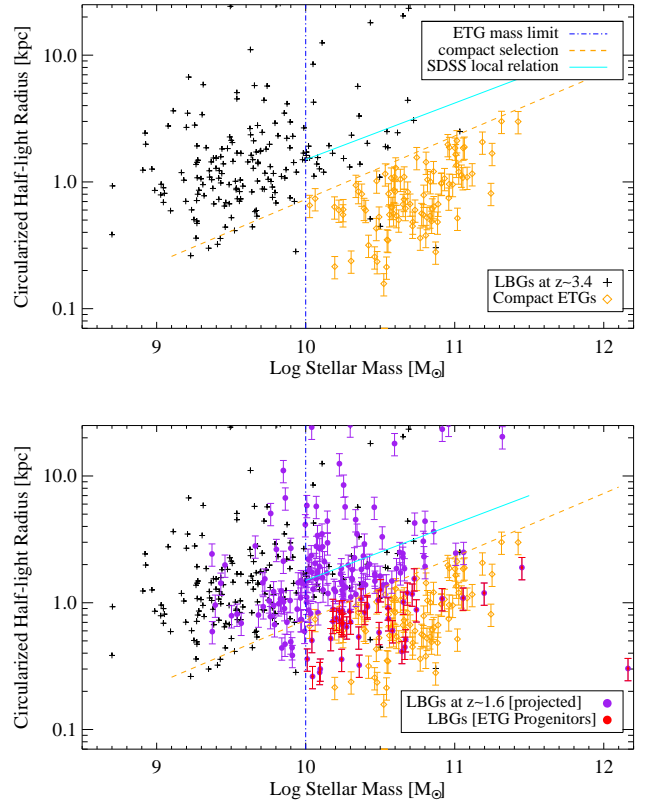}
\caption{Top panel: Observed size-mass relationship of LBGs (black crosses), compared with that of the compact ETG sample of \citet{Cassata2013}, along with ETG mass selection (blue dashed line) and compactness selection (orange dashed line). Cyan line indicates the local mass-size relation of ETGs from SDSS \citep{Cassata2013}. Bottom panel: Same as top panel,  with additional points indicating the projected mass-size distribution of  LBGs after they satisfy the condition of passivity, namely $log_{10}sSFR<-2$ Gyr$^{-1}$ (purple and red points). Stellar mass of LBGs increases according to the assumed SFH and no size evolution (i.e. each  projected purple and red point has a corresponding black cross at the same half-light radius. See text for details). Candidate ETG progenitors are selected from this sample (red) as those LBGs whose projected mass puts them in the compact ETG  selection window. }
\label{sizevsm}
\end{center}
\end{figure}

It is important to understand that we are not postulating that all galaxies
that satisfy the two general requirements above will evolve by growing their
stellar mass at essentially constant size. On average, galaxies evolve by
growing their stellar mass and enlarging their size, as shown by the existence
of a mass--size relationship and the overall size evolution of star--forming
galaxies \cite[e.g.][]{Ferguson2004, Hathi2008b, Nagy2011, KHHuang2013}. But not all galaxies must do so, or must all follow
the same mass--size growth path, as evidenced by the scatter in the mass--size
relationship itself.  Nor are we saying that those LBG that satisfy the two
points above quench their star formation after the epoch of observation and
appear passive (as per our definition) at $\langle z\rangle\sim 1.6$. What we
are saying is that if LBGs at $z\sim 3$ include progenitors of the compact ETGs at
$\langle z\rangle\sim 1.6$, then the morphological properties of the latter
imply that the former must grow in mass at essentially constant size, as well as have
stellar mass and SFR such that under general assumptions about their
star--formation history subsequent to the epoch of observation, they quench and
are passive at the epoch when the compact ETG are observed.

To identify such LBGs, if they exist, we must model the quenching of their star
formation activity. Thus, we need to assume 1) the time when they begin
quenching, and 2) how they quench, i.e. the form of the declining star
formation history. The quenching phase can start at any time after the epoch
of observation of the LBG, or even slightly before, since a galaxy in the
early phase of declining star formation would still be classified as a
``Lyman--break galaxy'' as long as this phase is not too advanced.
The details of star-formation history of galaxies during the quenching phase
are not known \citep[][]{KSLee2011}, and there are suggestions that
star-forming galaxies can have bursty and chaotic SFHs \citep{KSLee2012a}.
Therefore a simple function such as an exponential decay is almost certainly
an over--simplification, especially on short time scales. If the goal is to
model the formation of ETGs, however, all that is relevant is that the
star--formation rate overall decreases on a relatively long time scale, namely
long compared to the time scale of rapid bursts (i.e. a few $10^7$ yr), since
their red colors (at the time of observation), high masses, and very low sSFRs
imply that they must have been actively star forming at least 1 Gyr earlier \citep{Onodera2012}. Previous studies investigating the SFHs
of the compact ETGs, have supported this interpretation \citep{Cimatti2008,
  Saracco2011, Saracco2012, Kaviraj2012a, Daddi2005}

\begin{figure} [!t]
\begin{center}
\includegraphics[scale=0.4]{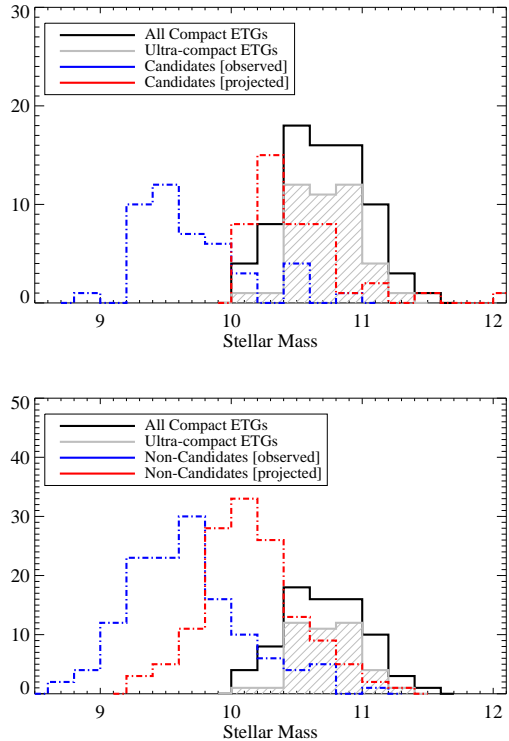}
\caption{The distribution of observed stellar mass and projected stellar mass
  (projected using the assumed SFH outlined in the text), for the candidate
  LBG sample (top panel) and non-candidate LBGs (bottom panel). For
  comparison, the mass distributions of all compact ETGs, and also the
  ultra-compact sub-sample are shown in black and grey, respectively}
\label{dmass_hist}
\end{center}
\end{figure}

It is important to keep in mind that the decaying exponential function that we
used to model the quenching phase of the LBG's star--forming activity has
nothing to do with the function adopted to describe their star--formation
history up to the time of the observation. The latter is used to model the
assembly of the galaxies that we observe and {\it only } to estimate their stellar mass at
the time of observation (recall that we tested both an exponentially declining
and a delayed increasing function when measuring the observed stellar mass, with similar results). To describe the SFH of the quenching phase, which begins at or after
the time of observation, we use only the exponentially declining SFH with $\tau$ = 100Myr as described in the Appendix, and this is used to estimate the additional stellar mass
produced after the time of observation and during the quenching phase (we do not make an attempt to incorporate mass loss from stars). This
stellar mass is then added to the mass of the LBG formed up to the time of
observation to derive the projected mass distribution of the candidate progenitors of
the compact ETGs and compare with the observed one for compact ETGs (see Figures \ref{sizevsm}
and \ref{dmass_hist}). In any case, the combination of the time when they start
quenching and the duration of the quenching phase must be such that by
redshift $\sim 1.6$ the galaxies would be observed as passive, according to
the criteria discussed above, and that the stellar mass distribution must
reproduce that of the compact ETGs.

Thus, once we assume the starting time of the quenching phase and the time
scale $\tau$ (e.g. the limiting case of the sound--crossing time, $\tau\approx
100$ Myr), from the measured (photometric) redshift, SFR and stellar mass of
the LBG, we can estimate the redshift at which the galaxy will satisfy the
condition for passivity, as well as its stellar mass at that
time. Additionally, from the measured size of the LBG, and under the
assumption that the galaxy evolves at constant size, we can estimate the
stellar density when the galaxy is passive and see whether or not it matches
the projected stellar density of the compact ETGs, i.e.  $\Sigma_{50}>3$x$10^{9}M_{\odot}$kpc$^{-2}$.

\begin{figure} [!t]
\begin{center}
\includegraphics[scale=0.4]{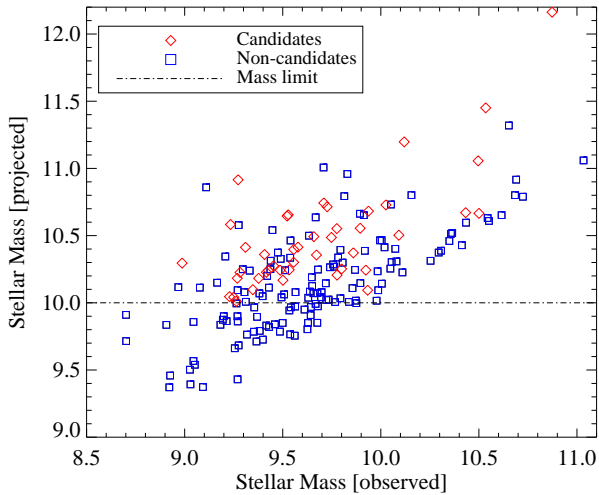}
\caption{The distribution of projected stellar mass (calculated
  with the assumed SFH outlined in the text), as a function of observed stellar mass,
  for the candidate LBG sample (red diamonds) and non-candidate LBGs (blue
  squares). The dot-dashed line indicates the ETG mass selection. Non-candidate LBGs which lie above this line are too extended to be selected as progenitors, and below this line non-candidates do not meet the mass selection. }
\label{dmass}
\end{center}
\end{figure}

We present our compact progenitor selection in Figure \ref{sizevsm}, which in
the top panel shows the mass-size relationship for the LBGs as observed, and
also the \citet{Cassata2013} ETG sample and its selection criteria.  In the
bottom panel, we additionally plot where the same LBGs will lie in the
mass-size diagram assuming their projected stellar mass, meaning, their
expected mass after they satisfy the condition of passivity, namely
$sSFR<-2$ Gyr$^{-1}$. Each LBG accumulates stellar mass according to its
observed SFR and assumed SFH. The distribution of final projected mass compared to the observed mass
is shown in Figure \ref{dmass}, and it should be noted that a large fraction
of the projected final mass of most galaxies must be extrapolated using the
observed SFR and assumed SFH. Those 44 LBGs with projected properties which meet
our compact ETG selection criteria are our sample of candidate compact ETG
progenitors. The rest of the LBGs (136), which end up either less massive or with lower stellar density, are non-candidates. We additionally note in this
Figure the existence of 11 LBGs from our sample who are {\it already} compact
in stellar density, as observed, at $z \sim 3$.

As might be expected, the candidate plausible progenitors tend to have higher
SFRs than the non-candidates. This trend can be seen in the SFR-M* relation
for the two samples, which is shown in Figure \ref{massvssfr}. There is one galaxy in the candidate sample for which we have estimated a rather large SFR ($\sim$10$^{4}$). We have investigated the SED of this galaxy for any signature of AGN, and found that it is not detected at 24$\mu$m, nor does it have an X-ray detection in the {\it Chandra} 4Ms image. This galaxy appears to simply be one of the redder galaxies in our candidate LBG sample, hence its high SFR. Even if our dereddening procedure overestimates the SFR of this object, which we think is quite likely, this galaxy will regardless end up in the candidate sample due to its high stellar mass and compact size already placing it in the candidate selection window. Its exclusion, or inclusion, does not significantly change the results presented in the following sections. We also note
that the candidates do not differ in their observed mass distribution from the non-candidates. This
can be further seen in their color-mass diagram for the samples considered
here, shown in Figure \ref{etgcolor}. We additionally present in Figure
\ref{etgcolor} the rest-frame U-V vs V-J color-color diagram, showing the
colors of the three samples of star-forming and passive galaxies. The color
distributions are important for understanding inherent differences in the two
samples of LBGs, as we will show in Section 4.

\section{Results}
\label{Results}
\subsection{Properties of compact progenitors}

\subsubsection{SEDs}

As mentioned in section 2, we have fit stellar population synthesis models
\citep{Bruzual2007} to each galaxy. 
The best fitting SED template to each galaxy, found as explained in Section 2, has been used to generate an average SED template for candidate and non-candidate samples.
We restrained ourselves to those LBGs in both samples
which have spectroscopically confirmed redshifts ($\sim 25 \%$ of all LBGs,
$\sim 44 \%$ of candidates, $\sim 18 \%$ of non-candidates), as this will
minimize the diluting effects of including objects with only photometric
redshifts which have larger uncertainties. We have also derived average
empirical SEDs directly from the observed photometry, k-corrected to common
restframe wavelength using the best fit templates. The errors on the average
empirical SED are estimated using the following procedure. With each galaxy's
observed photometry, we produce a Gaussian deviate of each photometric
measurement, given that point's photometric error. We then re-calculate the
average empirical SED. We do this procedure 10,000 times, and the standard
deviation of the average empirical SEDs of the gaussian deviates is the error
on the average.

\begin{figure} [!t]
\begin{center}
\includegraphics[scale=0.4]{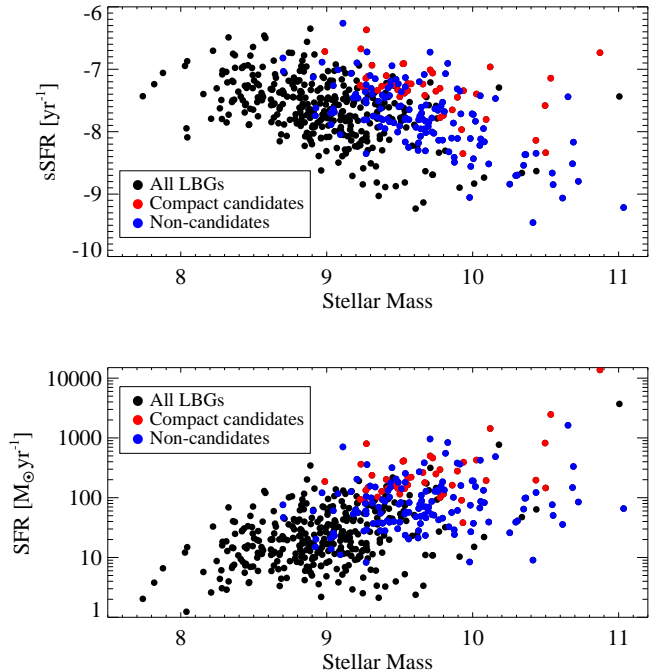}
\caption{Bottom Panel: SFR vs Mass distribution for candidate LBGs (red),
  non-candidate LBGs (blue) and compared with all LBGs selected from Figure 1
  (black) including H $>$ 25 objects. Top Panel: sSFR vs Mass for the same
  galaxies.}
\label{massvssfr}
\end{center}
\end{figure}
\begin{figure} [!t]
\begin{center}
\includegraphics[scale=0.4]{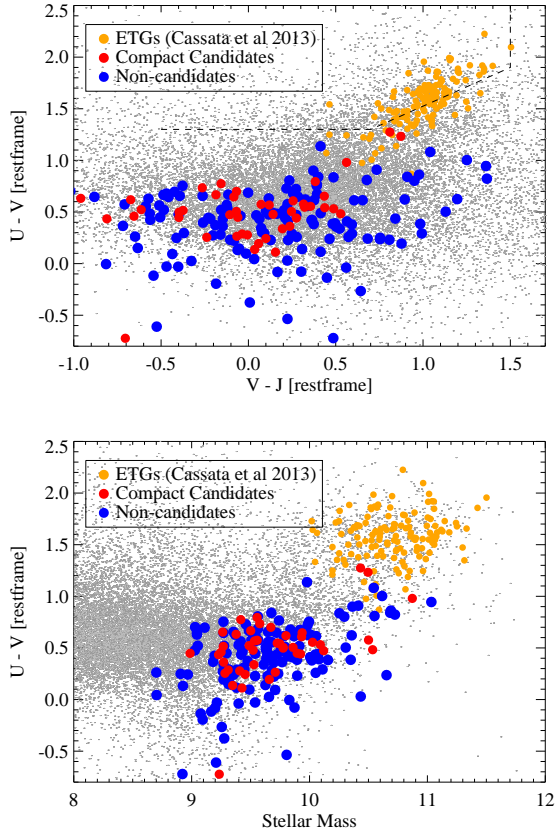}
\caption{Top panel: Restframe U-V vs V-J colors of the high-redshift ETG
  sample of \citet[][red points]{Cassata2013}, compared to those of the
  candidate (orange) and non-candidate (blue) LBG samples. Dashed line
  indicates the red sequence in UVJ color space as defined by
  \citet{RWilliams2011}.  Bottom panel: U-V color vs stellar mass diagram for
  the same galaxies. The candidate LBG sample tends to have redder U-V color
  distribution, while occupying a similar range in V-J to the
  non-candidates. Their stellar mass distributions are roughly similar.}
\label{etgcolor}
\end{center}
\end{figure}
\begin{figure*} [!!t] 
  \centering
  \includegraphics[scale=0.45]{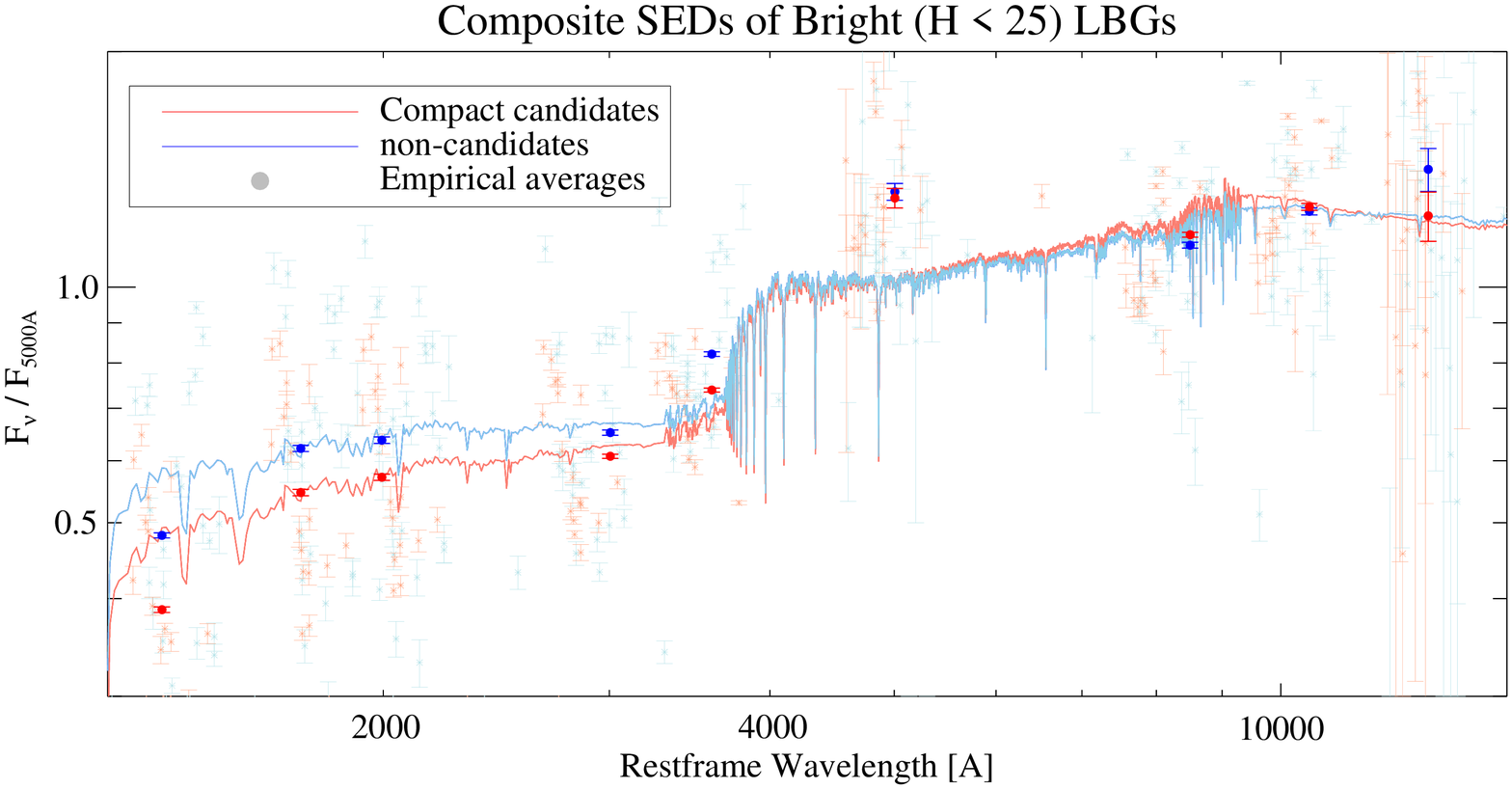}
  \includegraphics[scale=0.45]{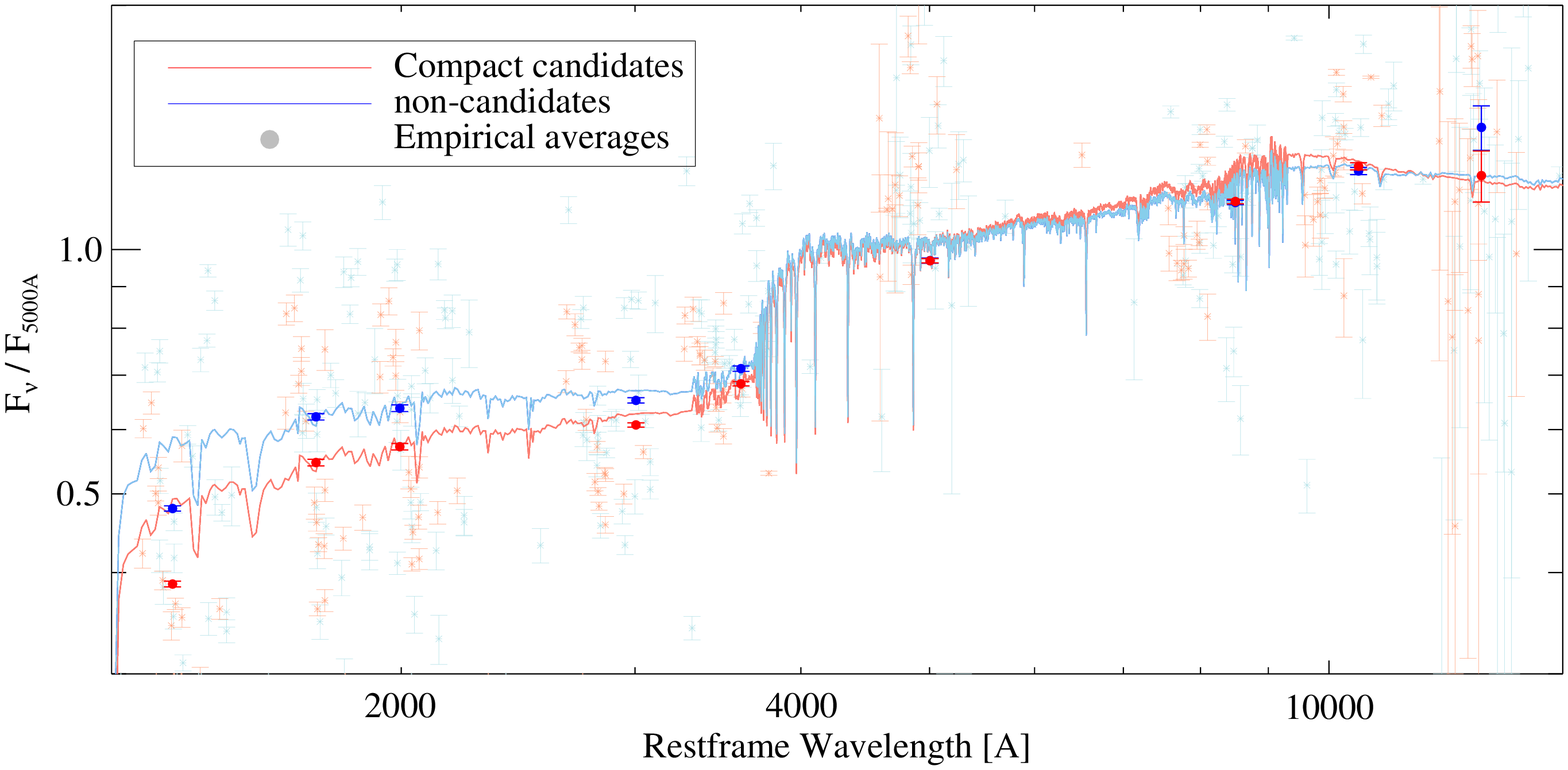}
\caption{Composite SED templates (lines) and composite of observed SEDs (circles) for the candidates (red) and non-candidates (blue) with spectroscopic redshifts (top panel). Observed photometry for the individual candidates and non-candidates with spectroscopic redshifts are also shown for candidates (salmon) and non-candidates (light blue). Bottom panel: same as
  the top panel, except photometric points which may be affected by emission
  lines have been excluded from the composite.}
\label{seds}
\end{figure*}

These two measures of the average SEDs for the two samples of spectroscopic
LBGs are shown in the top panel of Figure \ref{seds}. We have chosen to
normalize the average SEDs at $5000A$, the vicinity of the observed
$K_{s}$-band, so as to emphasize differences in the UV and optical parts of
the SED. At redshift $\sim 3$, there are four prominent emission lines
characteristic of star-forming galaxies which may enter the $K_{s}$ band (or
also the H-band, $H_{\alpha}$,$H_{\beta}$,OII, and OIII), and in fact in a
large fraction of our galaxies we do see an enhancement of the flux density in
the $K_{s}$ and/or H-band, relative to the best fitting SED
template. Therefore, we choose to normalize by the value of the best fitting
template at 5000A, rather than bias the normalization high by using the
observed flux density in the $K_{s}$ band.

The excess flux density of these contaminated photometric points with respect
to other photometric points, and also the best-fit SED, suggests the
photometry of some galaxies is affected by the presence of emission lines. We
investigate the extent to which these lines may affect the average SEDs in the
bottom panel of Figure \ref{seds} by repeating the analysis after removing the
individual affected photometric point from the galaxy's observed SED if one of
the four lines listed above enters any bandpass at greater than 1\% the
maximum transmittance of the band. The result of this test is shown in the
bottom panel of Figure \ref{seds}. It is clear that removing contaminated
photometry brings the points in the average empirical SED that are based on
the observed $K_{s}$ and H-band photometry into better agreement with the
average SED template. The difference between the SEDs of the compact
candidates and non-candidates is still clear when removing the affected
photometry.

The results shown in Figure \ref{seds} indicate that on average, the compact
candidates with spectroscopic redshifts have a redder restframe-UV SED than
the non-candidates, but an otherwise apparently identical optical one. There
are two explanations for a redder UV slope: an older population of stars (for
example, a more evolved burst of star-formation), or, a larger amount of dust
obscuration (on average). The effects of age or dust on galaxy SEDs are
generally degenerate and are notorious for confusing the measurement of
physical properties of galaxies. However, we argue here that the flatness and
consistency of the average SEDs of candidates and non-candidates red-ward of
4000A argues in favor of the interpretation that the difference is due to a
difference in average stellar ages (since peak of SF activity), and not
average dust properties.

\begin{figure} [!t]
\begin{center}
  \includegraphics[scale=0.4]{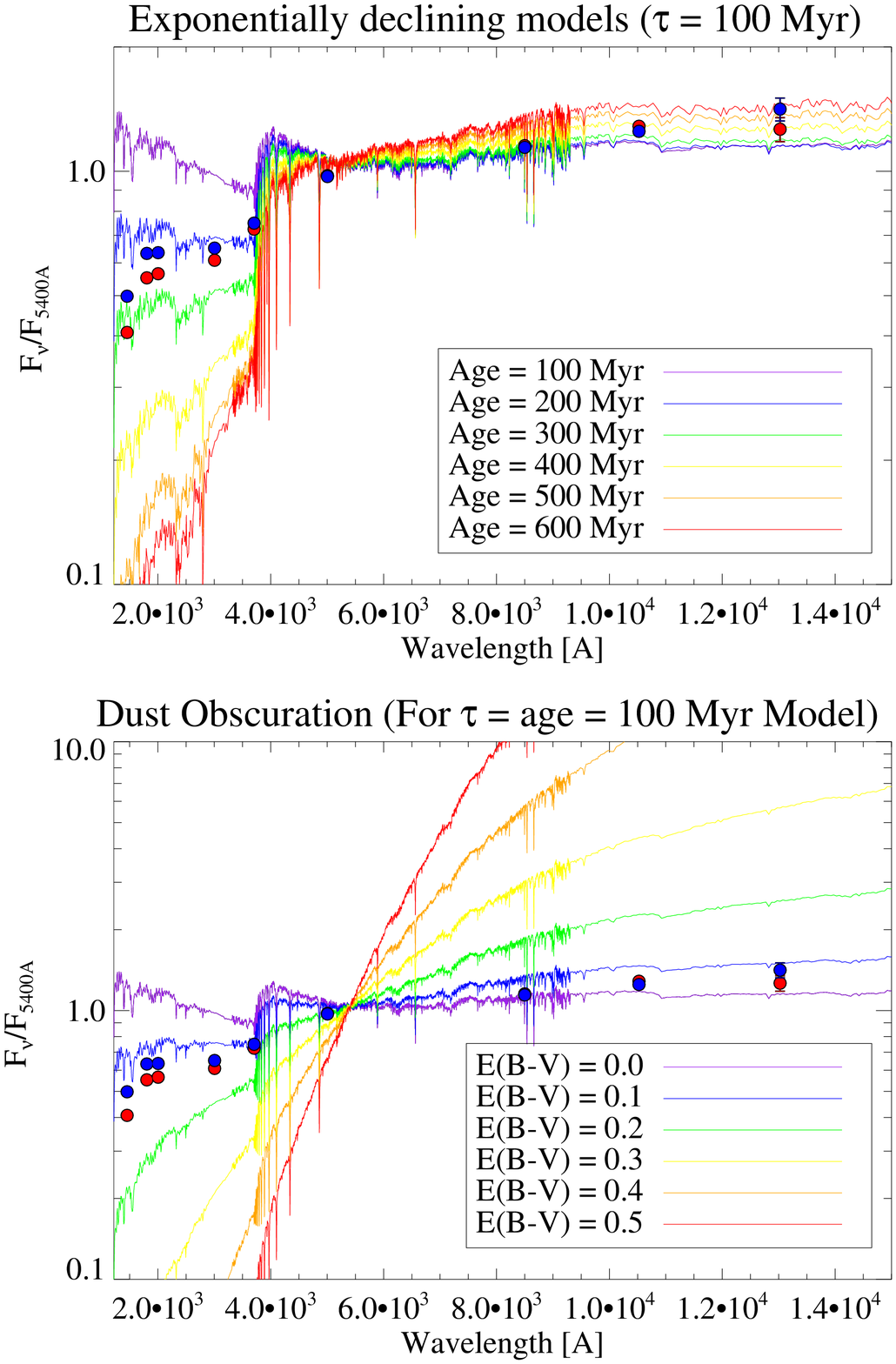}
  \includegraphics[scale=0.4]{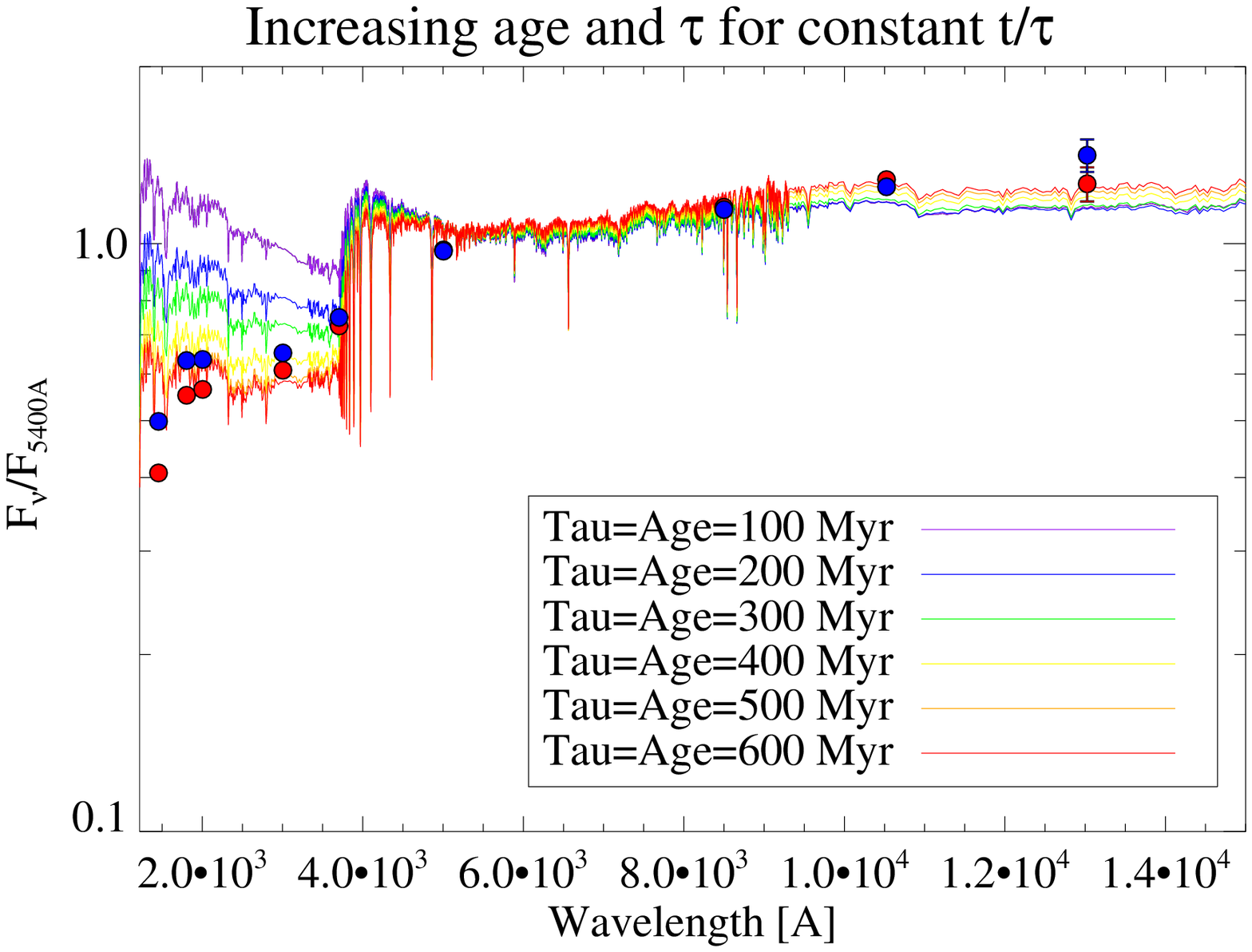}
\caption{Comparison of normalized SEDs for various exponentially declining
  SFHs with various age, tau, and dust. Top left: Model with $\tau = $100Myr,
  E(B-V) = 0, varying age of the burst. Bottom left: Model with $\tau =
  $100Myr and age = 100Myr, varying E(B-V). Bottom right: Model with E(B-V) =
  0, varying both age and $\tau$ such that the ratio remains
  constant. Composite SEDs of candidates (red) and non-candidates (blue) are
  included. Varying dust does not reproduce well the variation of the UV,
  while keeping the optical-NIR part flat.}
\label{modelseds}
\end{center}
\end{figure}

Figure \ref{modelseds} illustrates how the UV and optical parts of
\citet{BruzualCharlot2003} SED templates of star--forming galaxies vary with
varying age, varying dust obscuration, and varying both age and tau for
constant t/$\tau$ ($t/\tau \propto sSFR$). The model SEDs are normalized in
the same way as the observed data, and the averages for candidates and
non-candidates are included for comparison. Varying dust does not reproduce
well the difference in composite SEDs, because of the similarities in the
optical part of the SEDs of both candidate and non-candidate samples. Varying
age is a better description of the observed trends.

\begin{figure*} [!ht]
  \begin{center}
  \includegraphics[scale=0.4]{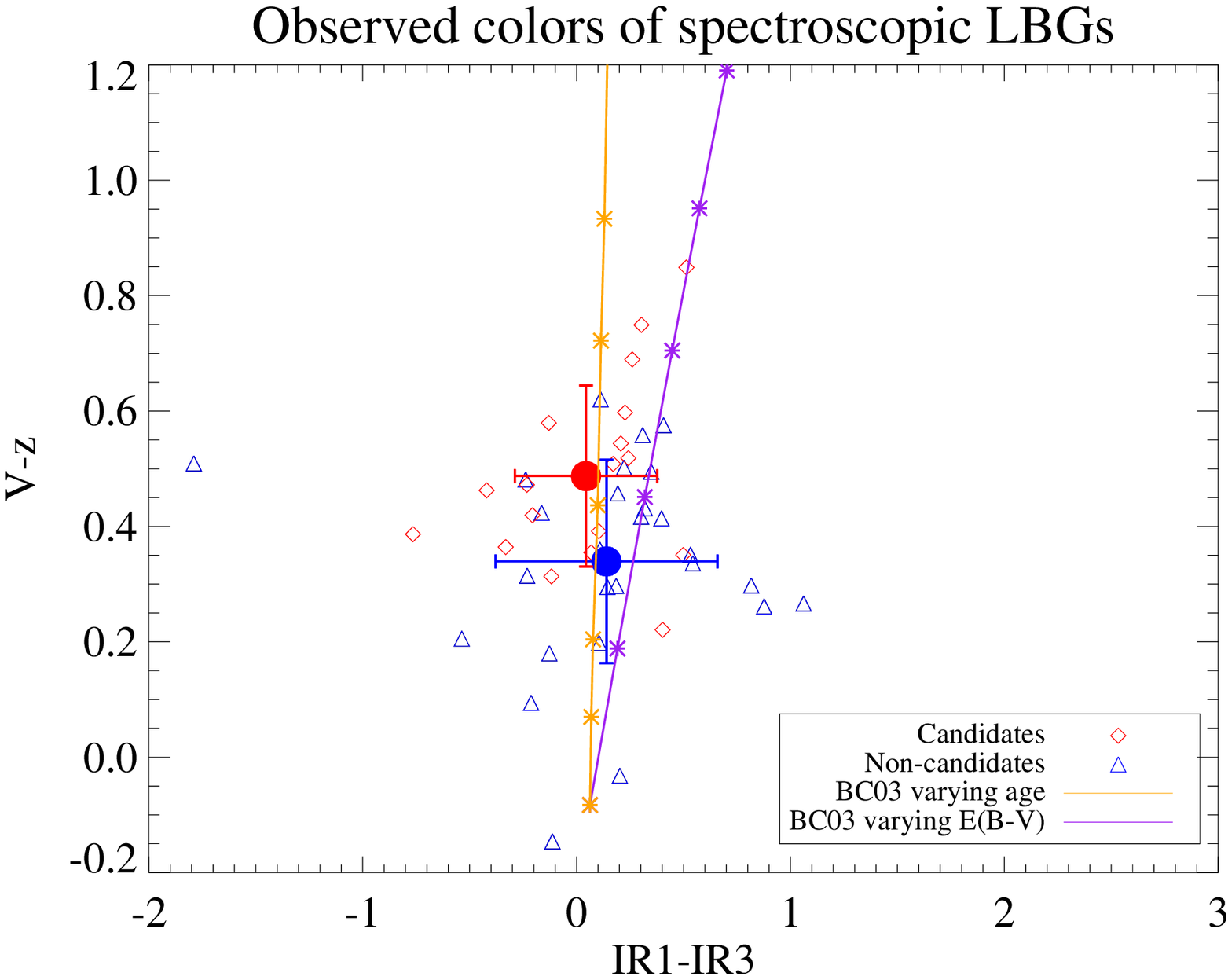}
  \includegraphics[scale=0.4]{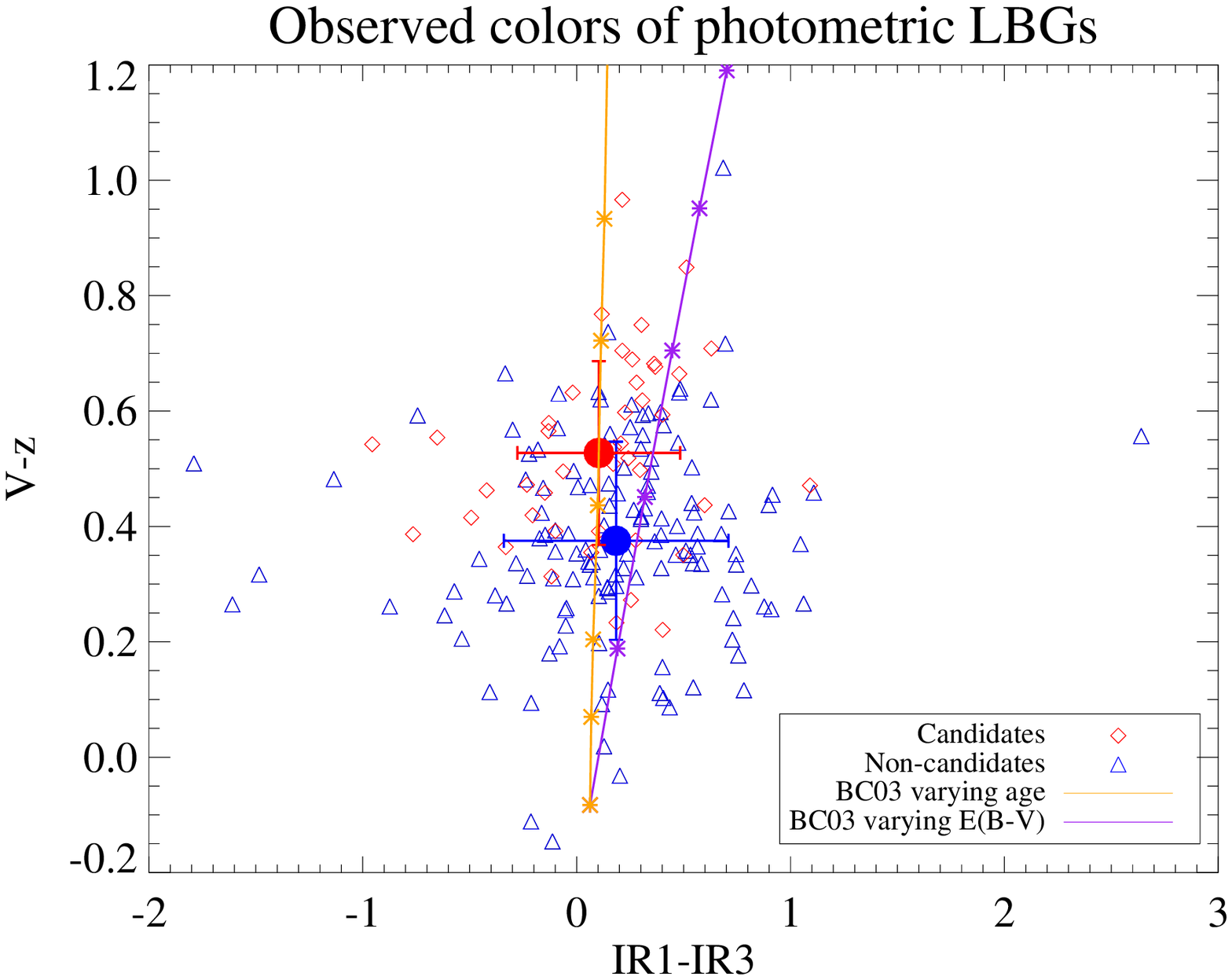}
\caption{Observed color-color diagrams for candidates (red) and non-candidates (blue), for
  the spectroscopic sample (left) and photometric sample (right). Large solid
  circles and their error bars represent the mean and standard deviations,
  respectively, of the candidates (red) and non-candidates (blue). Over-plotted
  are the colors of the \citet{BruzualCharlot2003} models from Figure
  \ref{modelseds}, with $\tau = 100$Myr at z$\sim3$, for varying age from
  100-700 Myr (orange) and varying E(B-V) from 0 to 0.7 (purple).}
\label{obscc}
\end{center}
\end{figure*}

A better look at the photometric differences between candidates and
non-candidates and how they may vary with trends in age and dust can be seen in Figure \ref{obscc}, which shows the observed V-z
and IRAC channels 1-3 colors of LBGs. This compares colors blue-ward of the
4000A break with colors red-ward, which are the key features of the differing
average SEDs. There is overlap in color distributions of the candidates and
non-candidate samples, but the mean of the candidates are offset red-ward in
V-z from the non-candidates, but not offset red-ward in IRAC colors. The same
color distribution is present among the LBGs which have photometric redshifts
(bottom panel), indicating this difference in color persists among the entire
LBG sample. Tracks of a sample \citet{BruzualCharlot2003} template SED with
varying age and one with varying dust obscuration show that varying age
changes V-z color but not IRAC colors, while varying dust changes both
colors. This is consistent with, although does not prove, a difference in the average stellar age (since
peak of the starburst) between the two samples, with the candidate ETG
progenitors appearing older than non-candidates. We note that similar
trends are also seen in the U-V vs V-J color-color diagram presented in Figure
\ref{etgcolor}, where candidates tend to be redder in U-V, but do not appear
redder in V-J color, than the non-candidates. Although the dispersion in colors between candidates and non-candidates in Figure \ref{obscc} overlap, we note that the difference in average SEDs are significant, as determined by the more robustly determined errorbars from the simulations presented in in Figure \ref{seds}.

\subsubsection{Infrared Properties}

Here we present further investigation of the potential contribution of dust to the redder restframe-UV SED of the candidates as compared to that of the non-candidates, using the {\it Spitzer}/MIPS 24$\mu$m  \citep[][Dickinson et al in prep]{Magnelli2011} and {\it Herschel}/PACS 100$\mu$m data \citep{Elbaz2011, Lutz2011}.
The 3$\sigma$ detection limit of the 24$\mu$m catalog is 20 $\mu$Jy, and for the 100$\mu$m catalog (based on prior information of the 24$\mu$m catalog) is 0.8 mJy \citep{Elbaz2011}. 
 Generally the LBGs have a low detection rate ($<8\%$) at both 24$\mu$m, and 100$\mu$m. Of the candidates, 4$\%$ are within 1 arc-second of a 24$\mu$m detection, and 2$\%$ are within 1 arc-second of a 100$\mu$m detection. 
 Of the non-candidates, 8$\%$ are within 1 arc-second of a 24$\mu$m detection, and 3$\%$ are within 1 arc-second of a 100$\mu$m detection.

Although the detection rate is low, we checked to see if these galaxies should have
infrared fluxes above the detection limits of those surveys, given the amount
of dust obscuration that can be inferred from their measured UV-corrected
SFRs. To see if our LBGs should be detected given this estimated amount of
dust-obscured SF, we first estimate the infrared luminosity
($L_{IR}$(total)=$L$(8-1000$\mu$m) from the amount of obscured SFR in each
galaxy. The obscured SFR is derived from the UV-corrected SFR and the
UV-uncorrected (i.e. measured directly from the restframe-UV SED) SFR. To
estimate the expected flux densities at 24 and 100-$\mu$m, we use the infrared
template of \citet{CharyElbaz2001} whose total infrared luminosity best
matches that estimated for each LBG, convolved with each bandpass. We find
that less than 3(2)$\%$ of the non-candidates are expected to be detected at
24(100)$\mu$m, and less than 7(7)$\%$ of the candidates would be detected. We
also test the inferred flux densities using the updated templates of \citet{Elbaz2011} for main sequence galaxies, and find comparable results. These are
generally consistent with our findings listed above for the actual number of
LBGs with infrared counterparts.

Since the overwhelming majority of these LBGs are below the detection limit of
the {\it Spitzer}/MIPS and {\it Herschel}/PACS surveys, we study the average
dust properties with a stacking analysis.
Because the detection rate within a 1 arc-second search radius is so low, we
adopt the following procedure for both wavelengths to carry out the
stacking. For objects which are within one arc-second of a detection, we use a
42x42 pixel image [50x50 arc-seconds] from the real observed 24$\mu$m or
100$\mu$m image. For objects which are not detected (more than 90$\%$) we use
a 42x42 pixel image from the residual map at 24$\mu$m or 100$\mu$m, where the
flux from formally detected objects has been removed using the PSF, following
the methods of \citet{Magnelli2011} at 24$\mu$m, and \citet{Elbaz2011} at
100$\mu$m. The residual maps thus includes low-level infrared emission from
non-detected sources, and also noise, but no flux from neighboring
detections. Using the residual map for non-detections minimizes the flux of
nearby bright, but unrelated, infrared sources contaminating the LBG stacked
flux. We then stack the images of the candidates and non-candidates using a
weighted averaging based on the rms maps in the case of the 24$\mu$m stack and
the weight maps in the case of the 100$\mu$m maps. Stacked fluxes are
determined by performing aperture photometry on the stacked images, and the
published aperture corrections from \citet{Engelbracht2007} and the {\it Herschel}/PACS technical documentation. The uncertainty on the stacked
fluxes for candidates and non-candidates are determined by the following
procedure. We repeat the above stacking for the same number of random
positions in the maps as galaxies in the candidate and non-candidate
samples. We generate 1000 sets of these random stacks, and use the standard
deviation of the stacked fluxes of these random positions as the uncertainties
of each sample's stacked flux.

\begin{figure*} [!!t]
\begin{center}
  \includegraphics[scale=0.4]{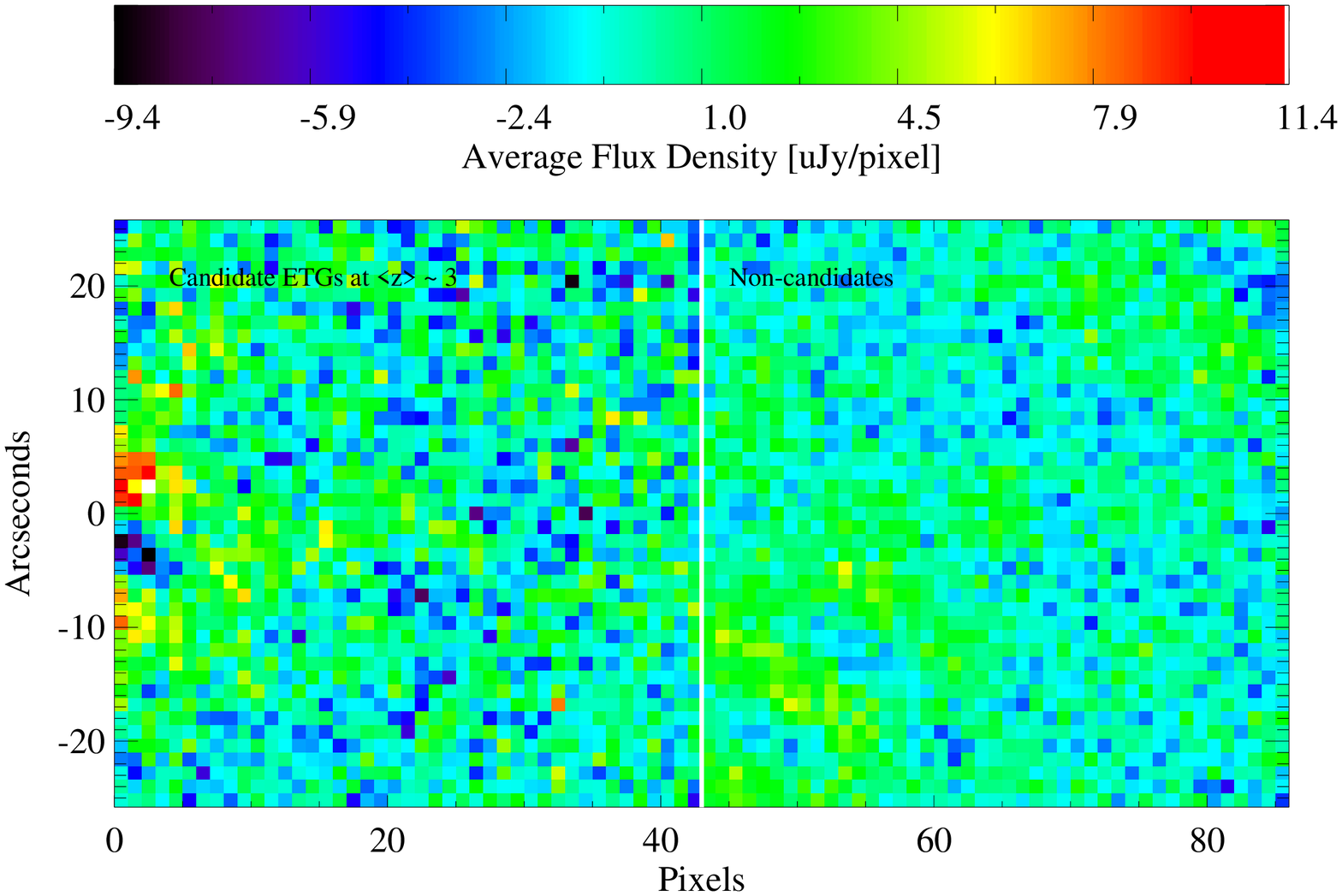}
  \includegraphics[scale=0.4]{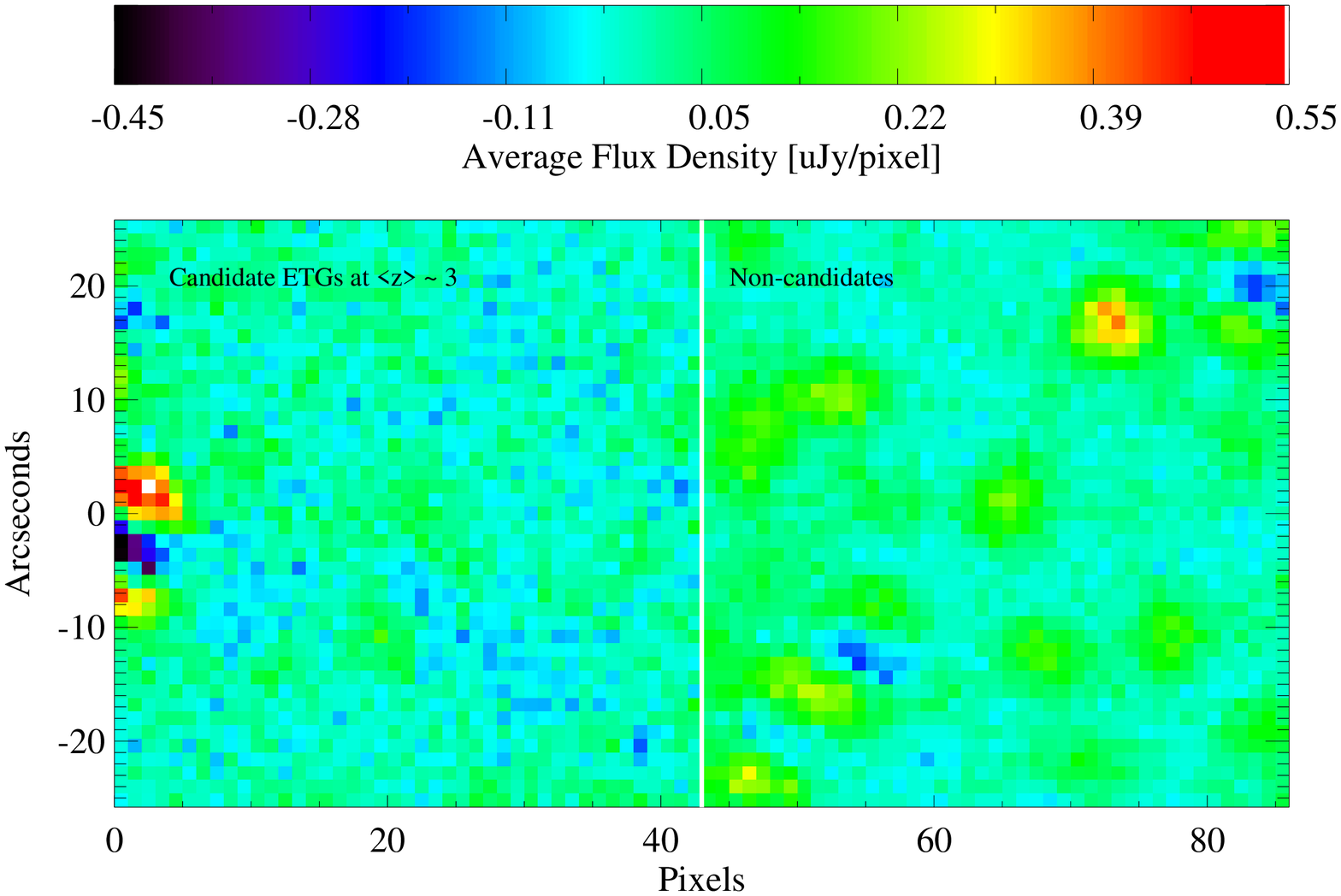}
\caption{Top panel: the stacked $Herschel$/PACS 100$\mu$m image, at the
  positions of the candidate LBGs (left) and non-candidates (right). No
  significant stacked emission is present in either sample. Bottom panel: The
  stacked $Spitzer$/MIPS 24$\mu$m image, at the positions of the candidate
  LBGs (left) and non-candidates (right). }
\label{stack100}
\end{center}
\end{figure*}

The results from the stacking analysis are shown in Figure \ref{stack100}. We
find that candidates and non-candidates have statistically indistinguishable,
and non-significant stacked flux at 100$\mu m$. For candidates we find
$F(100\mu m)_{stack} = 50\pm169\mu$Jy and for non-candidates we find $F(100\mu
m)_{stack} = -15\pm 87\mu$Jy. The candidates have no significant stacked
emission at 24$\mu$m, while the non-candidates do have some significant
stacked flux. For candidates we find $F(24\mu m)_{stack} = 3\pm3\mu$Jy, and
for non-candidates we find $F(24\mu m)_{stack} = 9\pm2\mu$Jy, an approximately
4$\sigma$ detection. This supports our hypothesis that the candidates are
redder because of older ages rather than dust, because the candidates do not
show evidence of higher dust emission, and in any case, the non-candidates on
average appear to have more dust emission than the candidates.

\subsubsection{ X-ray stacking}
Of particular importance to studies of star-forming and compact galaxies, a
class to which our candidate LBG sample belongs, is the contribution of an
active galactic nucleus (AGN). Some galaxies whose flux is dominated by that
of an AGN may in effect {\it appear} compact in terms of bolometric output,
but are not in fact compact in terms of their stellar density. Rather, they
may simply be out-shined by the AGN. X-ray detections amongst our two LBG
samples thus play an important role in understanding the contribution (if any)
to contamination in our sample from bolometrically dominating AGN.

To assess the fraction of AGN present among the candidate sample, relative to
the LBG sample as a whole, we have matched the LBG samples to x-ray detections
from the 4 Ms observations with the {\it Chandra} X-ray Observatory
\citep{Xue2011}. The overwhelming majority of the LBGs are not detected in the
4 Ms observations, with the exception of 3 sources from the non-candidate
sample. Similarly low detection rates for LBGs have been found by
\citet{Hathi2013}. Nevertheless, since the non-compact sample is larger than
the compact one we cannot infer anything about the X-ray emission frequency in
the two samples. We further assess differences in X-ray properties between the
two samples using stacking of the 4 Ms imaging. In both the candidate and
non-candidate samples, the X-ray stack showed no statistically significant
X-ray emission, with an upper limit on X-ray luminosity of 10$^{43}$
erg/sec. This result suggests that our candidate sample is not contaminated by
AGN any more than the non-candidates.

\subsection{The UV and Optical Morphology and Size of The LBG samples}

In this section, we study the morphology and size of the young and old stellar
populations of the galaxies in each sample. The aims of this analysis are 1)
to explore faint surface-brightness features \citep[e.g.][]{Hathi2008} and 2)
to study if there are morphological differences in the two samples that can
help us assess if their merging histories differ.

Since our objects are faint (H$<$25), and because we are interested in low
surface-brightness features which may be too faint to observe on an individual
galaxy basis, we study the average distribution of the stellar populations by
stacking the two samples. We stack the $HST$/WFC3 H band, which is
in rest-frame optical at $z\sim3$ and therefore probes the older stellar
population. We also stack the $HST$/ACS (z band) images, which is at the rest-frame UV and therefore probes the young stars and star-forming regions. To
produce the stacked images, we generate images in each band, in which we have
masked out any neighboring galaxy which is not $also$ selected as a U--band
dropout by our color selection. To mask the emission from these interlopers,
we use the corresponding segmentation map from the sextractor detection
process in each band. The images are then shifted so that each image is
centered at the position of the peak of the H-band emission. To perform the
stack, we do an inverse--variance weighted mean of each pixel, using the map
rms plus poisson noise to compute the weights. This weighting scheme ensures
that possible non-azimuthally symmetric low--surface brightness structures will be preserved in the final stack. We then measure average
structural properties of the galaxies using GALFIT, and make azimuthally
averaged light profiles using the IRAF function {\it ellipse}.

\begin{figure} [!t]
\begin{center}
\includegraphics[scale=0.4]{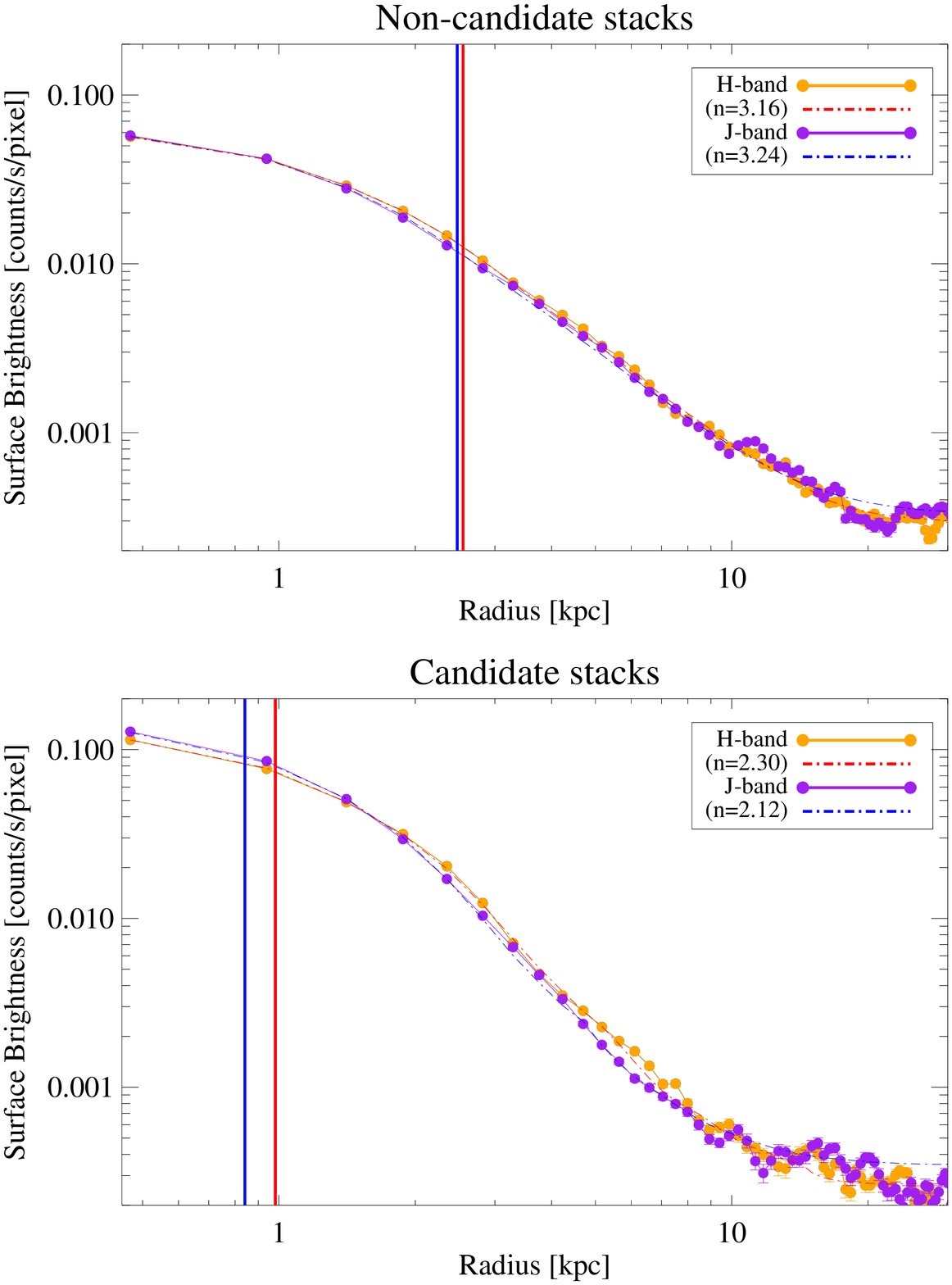}
\caption{J-band and H-band stacked light profiles for Non-candidates (top) and
  candidates (bottom). Points connected with solid lines indicate observed
  light profiles, and dot-dashed lines are the best-fit Sersic profiles as
  measured by GALFIT. Vertical lines indicate effective radius of each GALFIT
  model. Generally the structural properties in the two bands are very
  similar, suggesting that if emission lines from SF regions are entering the
  H-band bandpass, it is not significantly biasing our measure of the
  distribution of the old stellar populations.}
\label{jhprofile}
\end{center}
\end{figure}

As mentioned in section 4.1, given their redshift distribution, many of the
LBGs may have optical emission lines entering the H-band (primarily the
[OII]$\lambda 3727$, but also H-$\beta$), and this may bias the H-band light
distribution towards the star-forming regions and therefore would not trace
the older stars. The J-band is largely unaffected by these emission lines. We
therefore compare the H-band stacks and J-band stacks to check for
discrepancies, which may indicate contamination from this line emission. The
light profiles for the H-band and J-band, for candidates and non-candidates,
is shown in Figure \ref{jhprofile}. This figure shows that the H-band and
J-band stacks, on average, have very similar profiles for each sample, showing
that any contamination to the H-band from emission lines is negligible.

\begin{figure} [!t]
\begin{center}
\includegraphics[scale=0.35]{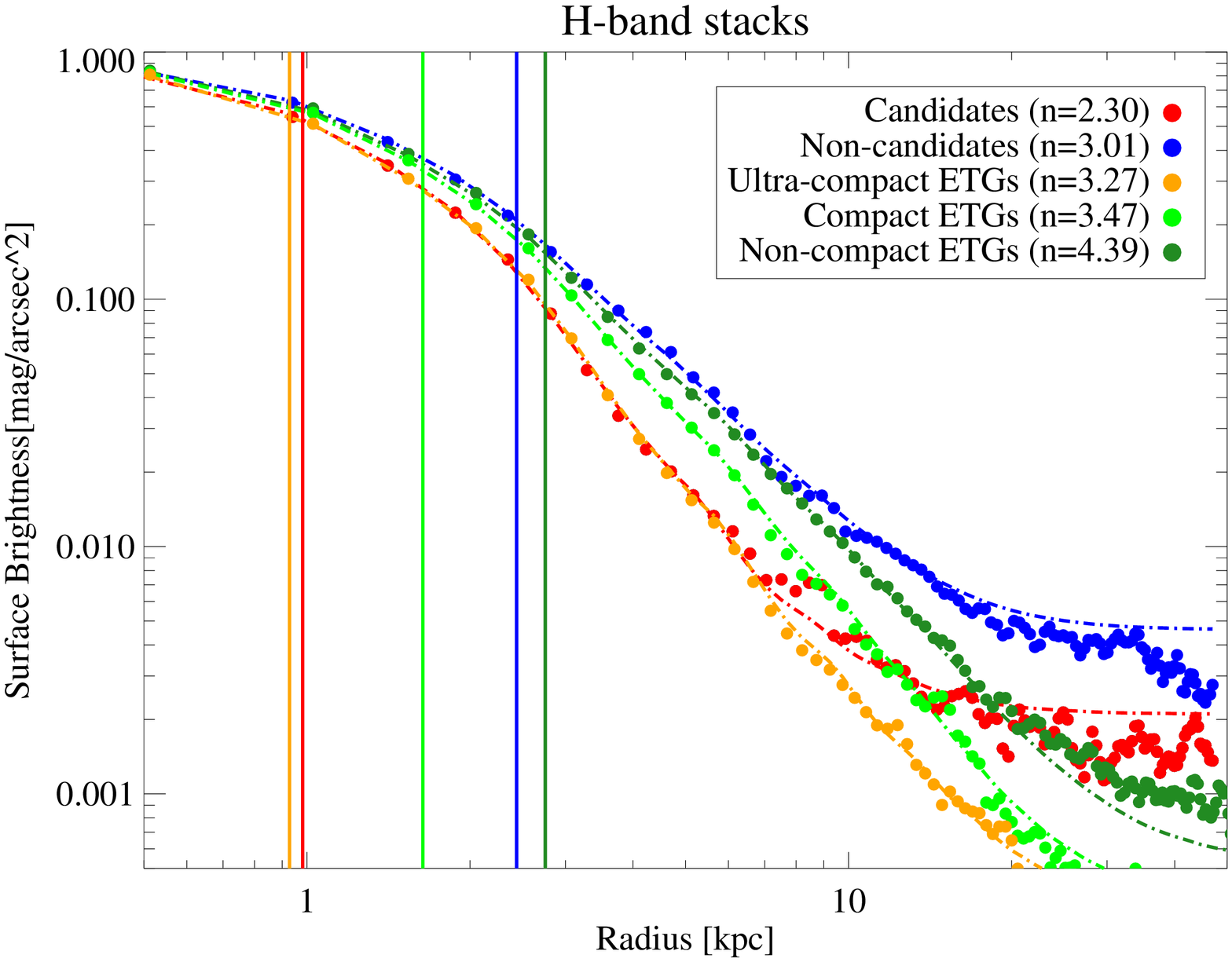}
\caption{Normalized light profiles (points) from the H-band stacks of
  candidates (red), non-candidates (blue) along with the those of the three
  classes of ETG samples of \citet{Cassata2013}. The three compactness classes
  are as defined in Section 3: non-compact (dark green), compact (green), and
  ultra-compact (orange). Best fitting sersic profiles to the stacks measured
  by GALFIT are dashed lines, and measured half-light radius represented by
  vertical line, and sersic index as indicated in the legend.}
\label{Hprofilesall}
\end{center}
\end{figure}

In Figure \ref{Hprofilesall}, we compare the average, peak-flux normalized H-band light profiles for
the candidates and non-candidates. We additionally repeat the stacking using
the \citet{Cassata2013} compact ETG sample between $1.2 < z < 2.8$, where we
mask out all neighboring galaxies prior to stacking.  This comparison between
compact candidates (red) and non-candidates (blue) in this figure essentially
reflects our selection criteria: candidates must be compact (thus have smaller
radii) and tend to be more star-forming, resulting in a higher peak surface
brightness than non-candidates. (Due to the flux normalization in Figure
\ref{Hprofilesall}, the peaks are coincident, but this fact can be seen from
the absolute difference between peak and normalized noise level, and is also
visible by comparing the un-normalized peak flux for candidates and
non-candidates in the two panels of Figure \ref{jhprofile}).  We include in
Figure \ref{Hprofilesall} the stacked average light profile of ultra-compact,
compact, and non-compact ETGs as defined in Section 3 from the
\citet{Cassata2013} sample, which compares the peak-flux-normalized shape of
the light profiles. The figure shows that not only do the ultra-compact ETG
sample and our LBG candidate sample have, on average, the same half--light
radius (due to the way they have been selected), but also nearly
identically steep light profile overall. This presentation with the peak-flux normalization highlights the actual similarity in steepness of the light profiles. It is important to note here that simply comparing Sersic parameters does not adequately highlight the similarity because of the covariance of half-light radius and sersic index. Finally, we note that the average light
profile of all the samples considered here shows no evidence of excess flux
above a sersic profile, a signature which could imply the presence of tidal
debris from recent merging.

\begin{figure} [!t]
\begin{center}
\includegraphics[scale=0.4]{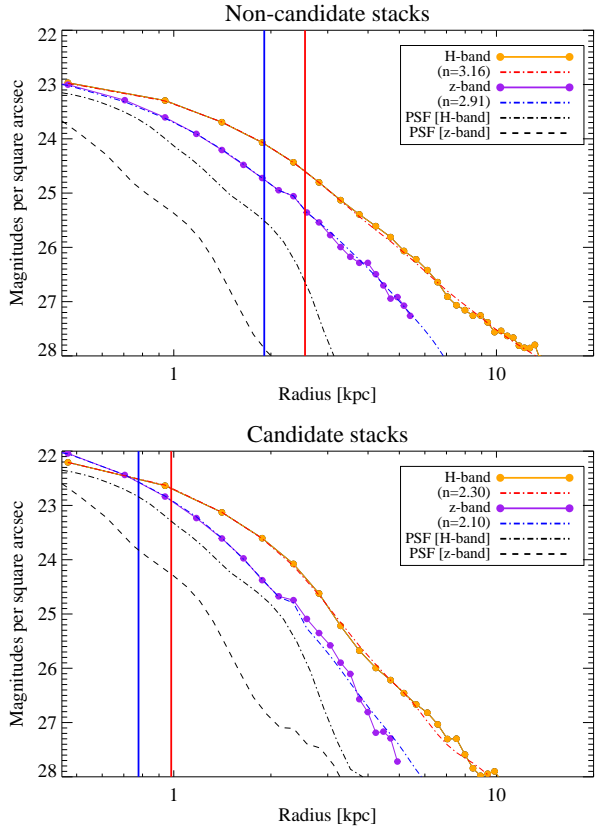}
\caption{z-band and H-band stacked light profiles for Non-candidates (top) and candidates (bottom). 
Points connected with solid lines indicate observed light profiles, and dot-dashed lines are the best-fit Sersic profiles as measured by GALFIT, convolved with the appropriate PSF in each band. Vertical lines indicate effective radius of each GALFIT model. 
}
\label{hzprofile}
\end{center}
\end{figure}

As mentioned previously, of particular importance to understanding the
mechanism by which these galaxies build up their stellar mass over time is the
relative {\it spatial} distribution of young and old stars. In Figure
\ref{hzprofile}, we compare the average stacked light profiles of the
restframe UV (z-band) with the restframe optical (H-band) for our candidates
and non-candidates. In this Figure, the differing PSFs have been fully taken into account by GALFIT when estimating the sersic parameters, and are plotted after convolution with their respective PSFs. In both candidates and non-candidates, it is clear that
the UV flux is more centrally concentrated than the optical flux, as indicated
by the fact that the UV half-light radius of each is smaller than that of the
optical, lending support to our assumption in section 3 that these galaxies
gain stellar mass through centralized, in situ star-formation, rather than in
the outskirts for example, and therefore may build the stellar masses of
galaxies at constant (or modestly increasing) size, consistent with the
evolutionary path we assume in Figure \ref{sizevsm} (for a similar discussion
on how galaxies grow in mass and size see also \citet{Ownsworth2012}, who,
however, do not consider the case of compact galaxies as we do here). 

\subsection{Number densities of candidates and real ETGs}

If major merging has not been a driving mechanism, then the co-moving number
densities of our candidate progenitors and that of the compact ETGs should be
comparable. In the literature,
co-moving number densities have been used as arguments favoring massive dusty
starbursts, such as ultra-luminous infrared galaxies (ULIRGs) and
sub-millimeter galaxies (SMGs), as being the progenitors of local massive
ellipticals in the centers of galaxy clusters \citep{Swinbank2006,
  Daddi2009}. These galaxies are relatively rare ($n \sim 10^{-5}$ Mpc$^{-3}$,
\citet{Scott2002, Chapman2005}) like the high-redshift ETGs ($n
\sim5.5(\pm0.8)$x$10^{-5}$ for non-compact ETGs and $n \sim
1.3(\pm0.1)$x$10^{-4}$ for compact ETGs, \citet{Cassata2013}). See Table
\ref{table1} for a comparison of number densities. Due to their extreme
nature, SMGs and ULIRGs may be capable of forming the requisite stellar mass
at high-redshift \citep{Lilly1999}, as well as have sufficient gas content to
produce compact remnants \citep{Tacconi2008, Tacconi2006}. (We will discuss
the similarities between SMGs and compact ETGs further in section 5.2.)
Evolutionary scenarios for SMGs can in principle be tested using galaxy
clustering \citep[e.g.][]{Hickox2012}, however current samples of
high-redshift compact ETGs is prohibitively small for this measure and
comparison.  Larger samples utilizing the full CANDELS survey data may contain
enough compact ETGs for this purpose in the future.

For now, co-moving number densities provide a first order consistency check. It
is important to keep in mind, however, that an accurate comparison of the
spatial abundances of the two populations would require not just quantifying
the effective amount of major merging that takes place both before and after
the star--forming galaxies quench, 
 but also the fact the LBG
selection does not recover all the star--forming galaxies at $z\sim 3$
 \citep[e.g.][]{Guo2012, Marchesini2010, Muzzin2013}. In any case, if our sample of compact candidates
truly contains the progenitors of compact ETGs, of which the majority grow
their stellar mass in--situ (i.e. steady growth independent of merging) and
evolve, by quenching,
 into compact ETGs, then the co-moving number densities
of compact candidate LBGs must at least be large enough to account for the
observed number densities of compact ETGs. We measure the co-moving number
density of our LBG samples by directly integrating the redshift distribution
to get the cosmological volume sampled by the galaxies:

\begin{eqnarray}
Volume = \frac{\int N(z) \frac{dV(z)}{dz} dz } {\int N(z) dz} \nonumber
\end{eqnarray}
where $N(z)$ is the redshift distribution of each LBG sample. We then use this to get the observed number density of LBGs in each sample:
\begin{eqnarray}
n = \frac{N_{tot}}{Volume} \nonumber \\
\end{eqnarray}

where $N_{tot}$ are the number of LBGs in each sample.
The co-moving number density of the \citet{Cassata2013} compact ETGs (stellar
density $>$ 3x$10^{9}$ $M_{\odot}$kpc$^{-2}$) is $1.3(\pm0.1)$x$10^{-4}$ Mpc$^{-3}$
between $1.2<z<2.8$. Co-moving number densities of other samples of compact
ETGs from other samples have similar values within this redshift and mass
range \citep{Barro2013, Patel2013}.

We find that our candidate compact ETG progenitors have a co-moving volume
density of $1.2$x$10^{-4}\pm 0.2$ Mpc$^{-3}$, consistent within the
uncertainty with the volume density of the \citet{Cassata2013} sample. More
specifically, the compact candidate sample can account for $\sim$92$\%$ of the
compact ETGs by number density found at lower redshifts. Although within the
uncertainty this can account for all detected compact ETGs, this does not
account for the destruction of compact galaxies through merging or
rejuvenation of star-formation between $z\sim3$ and $z\sim 1.6$.  In section
4.4, we will discuss a fraction of compact z$\sim$3 galaxies that we find are
missed by our LBG selection criteria, and how this fraction affects the above
estimate of number density.

\subsection{Compact progenitors missed by the LBG color selection}

We have chosen to use the LBG color selection for our sample in the above
analysis, because of its efficiency, lack of interloper contamination (with
U--band dropouts the only modest source of contamination is that by galactic
stars, which are easily identified in the {\it HST} images and removed) and it
is largely model independent \citep[see ][]{Giavalisco2002}. It also avoids the
potential bias due to the degeneracy between dust obscuration and age, which
could result in including galaxies with low specific star--formation rate
among the progenitors.

In any case, a more general search for candidate progenitors can be done 
using a sample selection based on photometric redshifts and subsequent SED
fitting to spectral libraries to derive stellar mass. In
fact, a fraction of $z\sim3$ galaxies with spectroscopic redshifts are missed
by our color selection, primarily due to the fact that they reside in crowded
fields where their TFIT photometry may be affected by nearby U--band detected
galaxies, causing them to be excluded from our sample of U--band dropouts.
Therefore, we briefly present again here our main results for candidates and
non-candidates selected in an {\it identical} way as the LBG candidates and
non-candidates, but this time from all galaxies with photometric or
spectroscopic redshifts between $3 < z < 4$. We will call this sample our
SED-selected sample.

Generally, the SED-selected samples are larger, with the number of
non-candidates increasing by $45\%$ and the number of candidates almost
doubling, with an increase by $95\%$. Not surprisingly, there is some
significant overlap between the LBG and SED-selected samples. Of the 136
LBG-selected non-candidates, 113 are in the SED-selected non-candidates. Of
the 44 LBG-selected candidates, 42 of them are in the SED-selected
candidates. (The two LBGs not present in the SED sample are galaxies with
photometric redshifts between 2 $<$z$<$ 3 discussed in section 3). Therefore
we stress these are not independent samples, and therefore this is not an
independent test of our results. Rather, adding the results from the
SED-selected sample serves to augment our analysis with an increased sample
size that is not biased by the fact that the LBG selection selects galaxies on
the basis of their UV emission, and therefore cannot be too dusty.  We
additionally note that while the LBG samples studied in section 4 have similar
24$\mu$m detection rates ($<8\%$), the SED-selected non-candidates remain with
a low-IR detection ($6\%$) and the SED-selected candidates jump to a $20\%$ IR
detection rate.

\begin{deluxetable*}{lccccl}

\tablecolumns{4}
\tablecaption{Volume densities of high-redshift galaxies\label{table1}}

\tablehead{\multicolumn{1}{c}{Galaxy Type} & \colhead{Notes} & \colhead{Co-Moving volume density\tablenotemark{a}} & \colhead{Reference}  }

\startdata

Sub-millimeter Galaxies & as observed & $1$ x $10^{-5}$  & e.g. \citet{Scott2002} & \\ 
Sub-millimeter Galaxies & incl. duty cycle & $1$-$3$ x $ 10^{-4}$ & \citet{Chapman2005},  & \\
 & & & \citet{Swinbank2006}  & \\
ULIRGs & as observed & $7.5$ x $10^{-5}$ & e.g. \citet{Magnelli2011} & \\ 
Ultra-compact ETGs & $1.2 < z < 2.8$ & $7.4(\pm1.1)$ x $10^{-5}$ & \citet{Cassata2013}  & \\ 
Compact ETGs & $1.2 < z < 2.8$ & $1.3(\pm0.1)$ x $10^{-4}$ & \citet{Cassata2013}  & \\ 
Non-compact ETGs & $1.2 < z < 2.8$ & $5.5(\pm0.9)$ x $10^{-5}$ & \citet{Cassata2013}  & \\ 
Compact candidate LBGs & $3 < z < 4$ & $1.2(\pm0.2)$ x $10^{-4}$ 
&  This study  & \\
Non-candidate LBGs & $3 < z < 4$ & $3.7(\pm0.3)$ x $10^{-4}$ 

 &  This study  & \\ 
All H $<$ 25 LBGs & $3 < z < 4$ & $5.1(\pm0.4)$ x $10^{-4}$  
&  This study  & \\ 
\enddata

\tablenotetext{a}{In units of [Mpc$^{-3}$]}
\end{deluxetable*}

\begin{figure*} [!!t] 
\begin{center}
\includegraphics[scale=0.45]{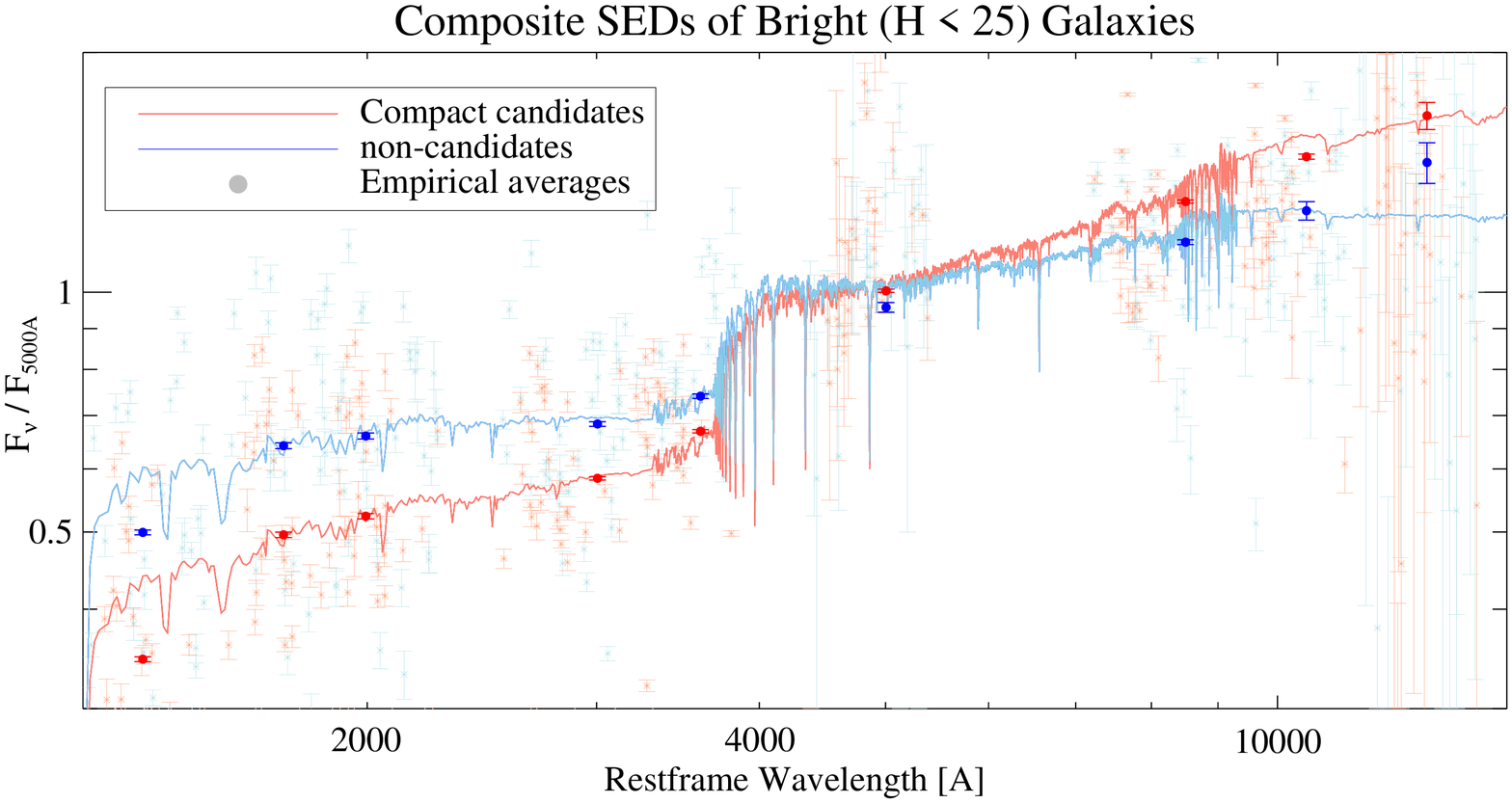}
\includegraphics[scale=0.45]{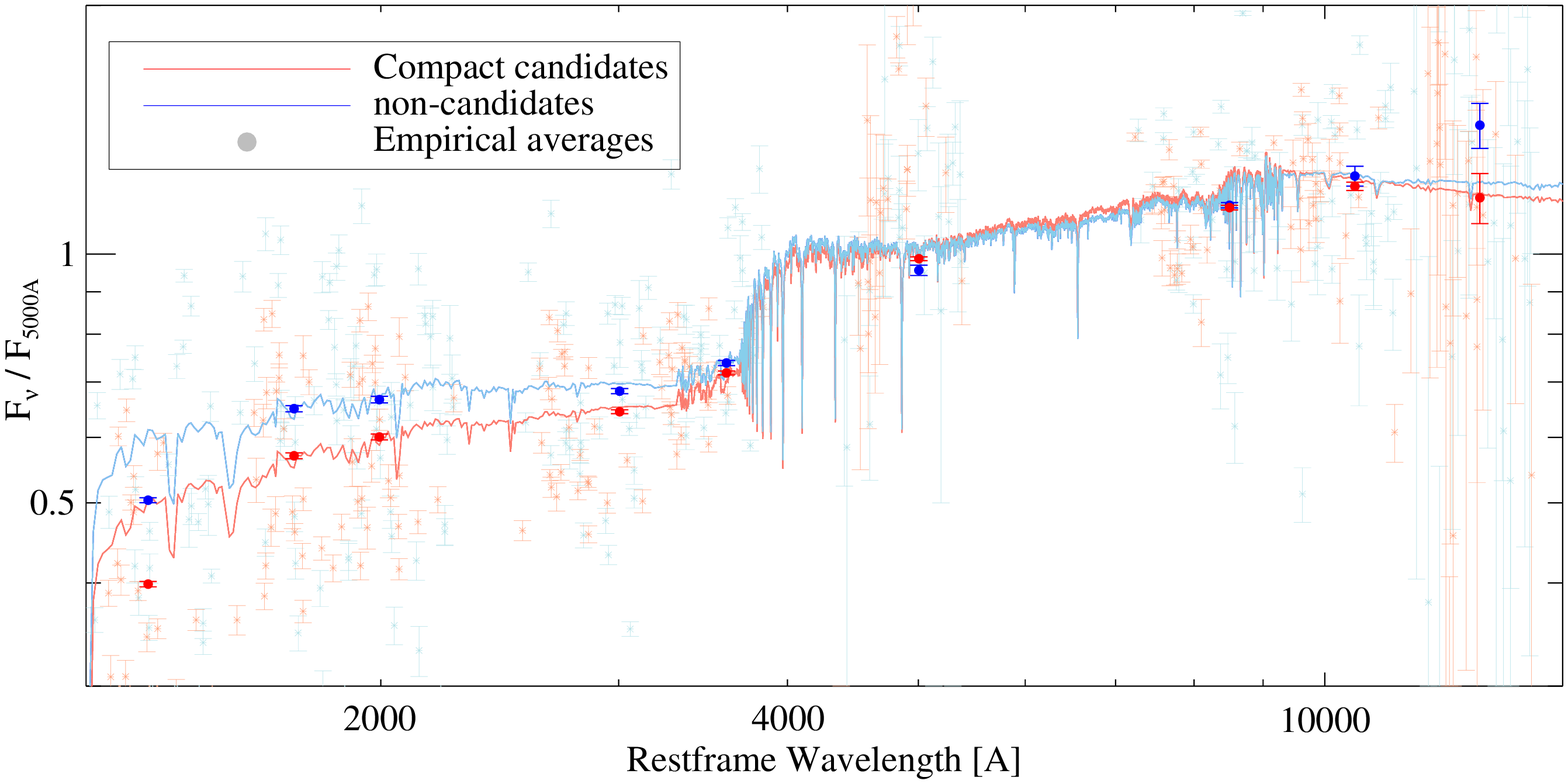}
\caption{Average SEDs, as calculated in Figure \ref{seds}, for candidates and
  non-candidates selected based on spectroscopic redshifts rather than LBG
  color selection. Top panel shows all candidates and non-candidates, and
  bottom panel shows the same sample with 24$\mu$m detected galaxies
  removed. Removing the small fraction of IR detected galaxies shows the same
  result as that from the LBG selection, indicating older ages among the
  candidates.}
\label{sedsamplesed}
\end{center}
\end{figure*}

We present in Figure \ref{sedsamplesed} the same analysis of average SEDs
presented in section 4.1, this time using the SED-selected candidate and
non-candidate samples. The top panel shows that candidates are redder, in both
the restframe-UV and the optical parts of the spectrum, suggesting a dustier
average SED. This is no doubt reflecting the increased fraction of IR
detections among the candidates using the SED-selection. However since the
number of IR detections in each sample is still a small fraction of the total,
we have repeated the analysis without including the IR-detected galaxies. The
bottom panel shows this comparison of non-IR detected candidates and
non-candidates, and shows the same signature of older stellar ages that was
seen for the LBG sample in Figure \ref{seds}. This figure, combined with the
results of section 4, are evidence that $3 < z < 4$ galaxies which are
selected to be compact in stellar density on average have a redder SED that is
best explained by an aging stellar burst. This supports our key result found for the LBGs in Section 4, namely that
there is a correlation between the compactness of galaxies, and the 
average age of their forming stellar populations, the sense being that more
compact galaxies have older forming populations, i.e. the starburst is older.

This SED-selected sample of compact candidates has a larger co-moving number
density of 2.6x10$^{-4} \pm 0.3$ Mpc$^{-3}$, compared to 1.2x10$^{-4}$
Mpc$^{-3}$ for the LBG-selected compact candidates. This number density of
SED-selected compact candidates (which includes $95\%$ of the LBG candidates)
is a factor of $\sim$2 larger than the compact ETGs.

\section{Discussion}

The key point of this paper is the identification of physically motivated
candidate progenitors of the massive, compact ETGs observed at $z\gtrsim
1.2$. Both individual images and very deep stacks show that the light profile
of the massive compact ETGs does not have any diffuse light in excess of their
extremely compact core, which is approximated by a steep Sersic profile with
typical parameters $n = 3.6$ and $r_e = 1.2$ kpc. Simulations show that
merging events rearrange the stellar mass profile of the merging partners in a
way that the profile of the merger remnant is more diffuse than that of the
initial partners \citep[e.g.][]{Lotz2010, Wuyts2010}. Dissipative processes in
a wet merger might channel gas to a nuclear region and produce a massive,
compact component in a starburst episode, however the pre-existing diffuse
component would remain visible and none is observed. This implies that
star--forming progenitors of the compact ETGs must be at least as compact
themselves, and thus we hypothesized that their progenitors may be found among compact star-forming galaxies at higher redshifts.

We looked for such progenitors and found some plausible candidates among LBG
at $z\sim 3$. The candidate progenitors have been chosen to have similar size
and morphology to the ETGs at $z\gtrsim 1.2$, and mass and star formation rate
such that after they quench their star--formation activity and would be
classified as ``passive'', e.g. according the definition given by
\citet{Cassata2011,Cassata2013}, their stellar mass and projected stellar
density are in line with the analogous properties of the ETGs. It is important
to understand that the selection of the candidate progenitors depends only
mildly on the details of the assumed time evolution of the quenching phase of
the LBGs; much more important parameters in determining if a given LBG is a
candidate progenitor or not are its physical size, its star--formation rate
and, to a lesser extent, its stellar mass at the time of observation, i.e. at
$z\gtrsim 3$. All of the observed $z\gtrsim 1.2$ massive passive galaxies with
stellar mass $M>10^{10}$ M$_{\odot}$ can be accounted for with progenitors
selected from LBG; we found that a factor of 2 more candidate progenitors are
found if more general selection criteria for star--forming galaxies are
adopted. With the addition of these new candidates, the evolutionary
constraints between these two populations is relaxed somewhat; up to half of
them may merge and increase size, rejuvenate their star-formation or fail to
quench, and there are still sufficient compact candidates to account for the
formation of compact ETGs. The recent study by \citet{Stefanon2013}, following a similar methodology to that presented here, found relatively few galaxies (5) which may be considered progenitors of compact ETGs, (although with higher mass limits of M$>10^{11}$ M$_{\odot}$ for ETGs, and initial z$>$3 star-forming samples with $10^{10.6}>M>10^{11}$). Their conclusion for these more massive samples differs from ours, in that a significant fraction of progenitors of these more massive compact ETGs must be created between 2 $<$ z $<$ 3 (they suggest through merging). Their sample is considered compact according to our criteria, and therefore likely overlaps with our sample on the massive end. We note that we do find a similar number (9) of SED-selected massive (M$>10^{10.6}$ M$_{\odot}$) galaxies that are already compact at z$>$3. To quantify any differences in the buildup of high-redshift ETGs as a function of stellar mass, for example if merging must contribute to the massive end as suggested by \citet{Stefanon2013}, relative to the more general ETG samples studied here where merging is not required, it will be necessary to study larger samples including other CANDELS fields in the future.

We subsequently studied the properties of the candidate progenitors and
compared them to those of the non-candidate LBG. The most remarkable
difference is that the candidate progenitors have significantly redder
rest--frame far--UV colors than the non-candidates but essentially identical
optical SED. The mid--IR properties of both types of galaxies show that larger
dust obscuration of the candidates vs. the non candidates is unlikely to be
responsible for the difference. If anything, the non--candidates have larger
dust luminosity given their activity of star formation. This forces us to
conclude that the redder UV is explained by an older burst (or period) of star
formation, namely the phase of star formation of the candidates has progressed
more toward quenching it. This is consistent with the general finding that a
compact stellar morphology is the best predictor of passivity, i.e. early
star--formation quenching, among massive galaxies at $z>1$
\citep[e.g.][]{Bell2012}.

 The fact that position of the candidate progenitors on the main sequence
  is higher than that of the non--candidates is particularly interesting in the context of the discussion of the main sequence of galaxies presented in \citet{Renzini2009}. The arguments presented there suggest that galaxies on the main sequence with above average SFR must quench rapidly and early, resulting in a generally shorter lifetime of star-formation. It is therefore interesting that not only do we find that our candidate sample follows such an elevated distribution of star-formation rates on the main sequence relative to the non-candidates, as seen in Figure \ref{massvssfr}, but we also find that they appear to show signs their star-formation will shut down. Qualitatively speaking, the candidates are consistent with this scenario outlined in \citet{Renzini2009} for a more rapid evolution for galaxies which follow the main sequence with higher SFRs.

The rest--frame UV and optical morphology of both the ETGs and of
the LBG candidate progenitors show the lack of any diffuse component in excess
of the very compact main body of the galaxies. This is observed both in
individual galaxies and in deep stacks. Additionally, LBG candidate progenitors and ultra-compact ETGs show similarly steep profiles (see Figure \ref{Hprofilesall}). The fact that star formation is
observed to take place only in very compact regions in the LBG progenitor
candidates and that no ``halo'' stars (down to a limit of $\sim$29 magnitudes per square arcsec), 
i.e. no stellar component in excess of the steep light profile of the compact
body of the galaxies, are observed in the passive ones is consistent with our
suggestion that the growth of stellar mass in these galaxies takes place at
essentially constant size, i.e. the stellar density increases with the stellar
mass. In turn this is indicative that star formation takes place in these
systems via a highly dissipative accretion of gas. We speculate in the following sections that this
might be consistent with some ideas on how cold accretion proceeds in massive
galaxies \citep[e.g.][]{Dekel2009b, Dekel2009, Oser2010, Sales2012, Johansson2012}.

\subsection{In-situ star-formation from accretionary mechanisms}

The existence and the spatial abundance of compact candidate progenitors shows
that there are sufficient ordinary star-forming galaxies with small size
and large stellar mass and star--formation rate that can evolve through in
situ star-formation to form compact ETGs, without having to invoke special
mechanisms that in a relatively short timescale can both quench a star-forming
galaxy, while at the same time change its morphology into a compact one,
e.g. gas-rich mergers with differential dust obscuration to hide the extended
halos \citep[see][]{Wuyts2010}, to explain the emergence of compact
ETGs. Rather, simple secular evolution through in situ star-formation is
sufficient, provided a quenching mechanism exists that can sufficiently remove
the cold gas supply rapidly from these galaxies.  We discuss various arguments
in favor of this scenario below.

\subsection{Comparison to Ultra-luminous infrared galaxies and submillimeter galaxies } 

ULIRGs and SMGs are often identified as the progenitors of local ETGs, as well as the compact ETGs at
high-redshift, based on number densities \citep{Daddi2009, Swinbank2006},
velocity-size relations \citep{Bouche2007}, consistency with the Faber-Jackson
relation \citep{FaberJackson1976} for local ellipticals \citep{Swinbank2006},
and clustering \citep{Blain2004, Brodwin2008, Hickox2012} \citep[although
  see][for the limitations of SMG clustering measurements]{Williams2011}. The
connection between these IR-luminous galaxies and ETGs is based primarily on
the high SFRs and stellar masses that are typical of ULIRGs and SMGs, which
are capable of producing the requisite stellar mass of compact ETGs in short
timescales. Since these galaxies are commonly believed to be the result of
major mergers, the quenching of star-formation and morphological
transformations can be easily tied to the merger. However, studying the
stellar distributions of these galaxies is difficult due to their large dust
obscurations. This means only small numbers of SMGs have been mapped with HST
at near-IR wavelengths to measure the spatial extent of the bulk of the
stellar mass. The results from near-IR mapping indicates the morphologies of
SMGs are rather heterogeneous, including SMGs that are large and irregular
with multiple components, \citep{Smail2004, Swinbank2006, Tacconi2008}, dominant components which are disk-like \citep{Targett2011, Targett2012}, and
are on average larger in size than compact ETGs \citep{Swinbank2010, Mosleh2011,
  Targett2011, Targett2012, Bussmann2012}. Similar results have been found for
ULIRGs \citep{Kartaltepe2012}.  While a large fraction of these objects appear
to be gas-rich mergers \citep[][]{Tacconi2008, Kartaltepe2012}, collectively
these studies suggest SMGs and ULIRGs display a large variety of morphologies
and sizes, and as a population are not similar in morphology to compact ETGs.
While the SMG or ULIRG merger phase is likely a plausible avenue for quenching
massive galaxies \citep[e.g.][]{Hopkins2006}, 
its unclear how the average SMG can decrease in half-light radius by a factor
of 2 or more after the star-formation has been quenched in order to form a
compact ETG with the average properties of the \citet{Cassata2011, Cassata2013}  samples.

\subsection{ Formation of compact star-forming galaxies}

The more ordinary LBGs are generally smaller in radius compared to
star--forming galaxies at lower redshift with similar mass \citep{Ferguson2004, Nagy2011, Law2012}, but are larger than the compact
ETGs, as seen in Figure \ref{sizevsm}.  Despite this, our sample of compact
candidate progenitors make up the low end of the size distribution of $z\sim
3$ LBGs. How do such small but massive galaxies form?

According to major merger simulations \citep[see e.g.][]{Hopkins2008a,
  Wuyts2010}, with mass ratio from unity to $1:10$, compact star-forming
remnants may result from merging of galaxies with very large gas fractions,
e.g. larger than $\sim 40$\%. A compact remnant is formed when a sizable mass
of new stars is formed at the center of the new structure by the highly
dissipative gas. However, an extended stellar halo made by the older stellar
populations will remain. Partners that are compact at the onset of the merging
event, as well as a continuous accretion of gas as the merging progresses,
increase the mass fraction of the remnant in the compact structure relative to
that in the halo. But according to the simulations, however, a sizable
fraction of the stellar mass of the remnant will be found in the halo, in
general disagreement with observations of compact ETGs (some have resorted to
hide the extended halo with dust obscuration to bring the simulations in
better agreement with the observations \citep[see e.g.][]{Wuyts2010}).
In a simplified sense, gas--rich mergers produce two spatially segregated
stellar populations: a centralized starburst, embedded in an extended older
population.

In fact, a key feature of gas rich merger remnants seen generally among
simulations is the need for a two-component fit to the light profile: a
centrally steeper one to account for the starburst component and a shallower
more extended one to account for the remnant stellar component
\citep{Hopkins2008a, Wuyts2010, Bournaud2011}.
When fitting single sersic profiles to simulated gas-rich mergers,
\citet{Wuyts2010} find that the segregation of older and younger stellar
populations, primarily driven by the combination of light excess of the
dispersed older stellar populations at large radii and central cusp from the
starburst, causes the sersic fits to be better approximated with large values
of sersic index, $n$. As a result, they find the following distribution of
sersic indices and effective radii from gas-rich merger simulations: Merger
remnants are best fit by cuspy high sersic index fits ($n >$ 10), along with
large radii (driven by the extended component) too large to be consistent with
compact ETGs \citep[][their Figure 12b]{Wuyts2010}. Compared with observed
properties of the ETG sample of \citet{vanDokkum2008}, there appears to be
some disagreement between the observed measures, and those measured from
gas-rich merger simulations.

\begin{figure} [!t]
\begin{center}
\includegraphics[scale=0.42]{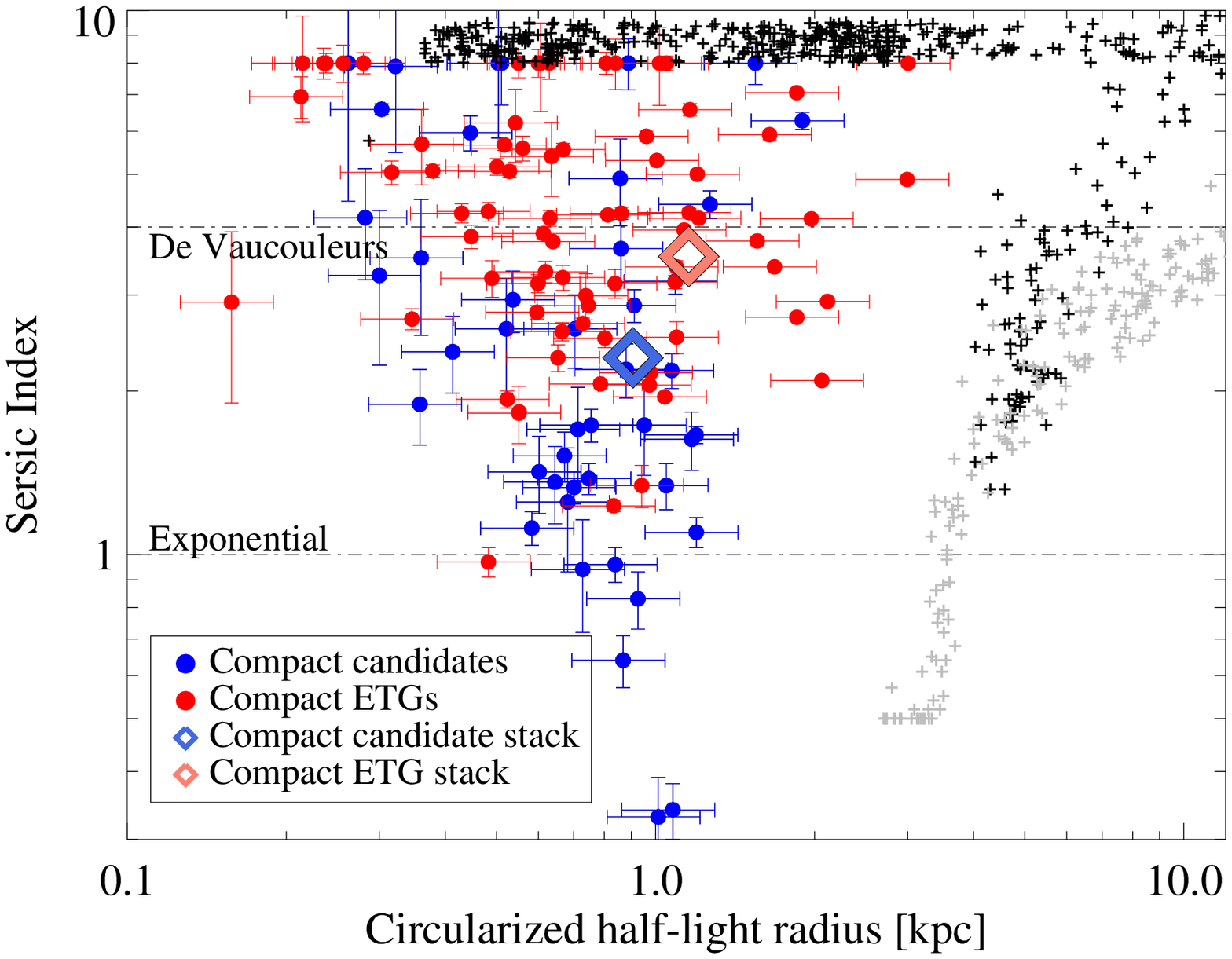}
\includegraphics[scale=0.42]{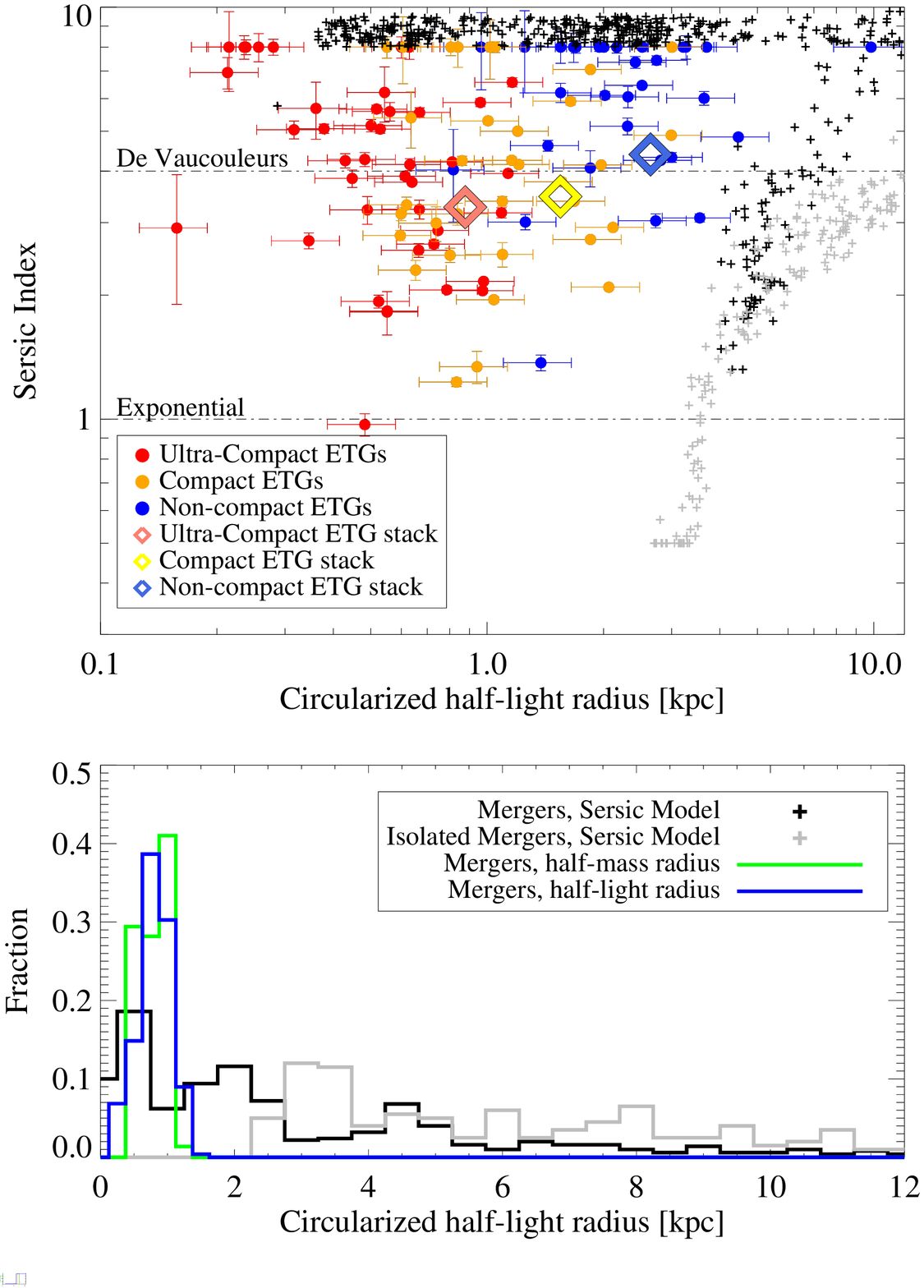}
\caption{Top Panel: Distribution of sersic index vs effective radius for
  candidates (blue) and compact ETGS (red). Measured values from the stacked
  images are shown in diamonds. Black and grey crosses represent the
  measurements of simulated gas-rich merger remnants and isolated mergers
  (isolated in the sense that the simulations do not include continuous gas
  inflow), respectively, from \citet{Wuyts2010}, their Figure 12b].  Middle
Panel: Same plot with the \citet{Cassata2013} sample, split according to the
compactness criteria.  Bottom Panel: The distribution of effective radii of
the \citet{Wuyts2010} models presented above (black and grey), as modeled by
sersic profiles, along with direct measures of half-mass radii, half-light
radii of the stellar component, and half-light radii after incorporating dust
attenuation.  }
\label{nvsre}
\end{center}
\end{figure}

For comparison, we have constructed a similar plot (Figure \ref{nvsre})
showing the distribution of sersic indices, as a function of effective radius
for the candidate sample and also the ETGs of \citet{Cassata2013}. We
additionally plot the measured properties from the H-band stacks for the
candidate sample and compact ETGs shown in Figure \ref{Hprofilesall}, as the
stacks are more sensitive to the presence (if any) of extended low-surface
brightness structures that would increase effective radius. These measures of
the light profiles of compact ETGs represent an improvement in sample size,
and in the case of the stacks, sensitivity to deviations from a sersic profile
and low surface-brightness features, relative to the comparison made with
\citet{vanDokkum2008}.

The distribution of our measures also do not coincide with the simulated
gas-rich merger measures, in that our galaxies are well-fit by single sersic
profiles, and do not show cuspy centers or extended light at large radii.  We
also see no evidence of these features in any of our stacks, with all light
profiles being well fit by single sersic profiles with $n \sim $2-4, down to
the noise limit in our stacks, and in general agreement with the distribution
of measurements from individual galaxies. To the extent that the simulations
represent real gas-rich mergers, our data of compact ETGs and compact
candidates (including both individual sersic fits as well as those of the
stacks) appear inconsistent with the simulations.  We note that the
distribution of radii of the models in the top and middle panels of Figure
\ref{nvsre} is mainly driven by the fact that sersic models are poor fits to
the merger simulations. However, the actual simulated half-mass and half-light
radii, without any assumption on the shape of the light profile, are still in
agreement with the observations. This can be seen more clearly in the bottom
panel, which shows that typical half-mass and half-light radii of the
simulations are of order $\sim$1 kpc. Therefore, we cannot rule out gas-rich
merging based on this analysis.

\begin{figure*} [!t]
\begin{center}
\includegraphics[scale=0.4]{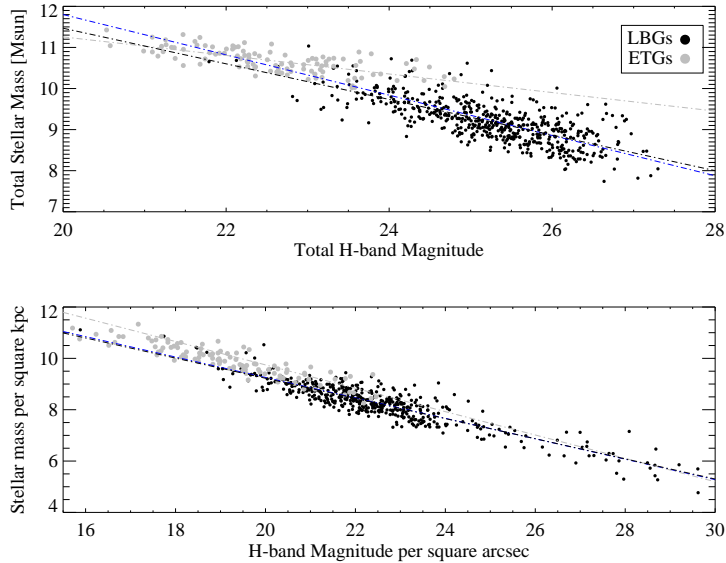}
\caption{Top panel: The correlation between the measured total H-band magnitude and total stellar mass for LBGs (black) and ETGs (grey). Linear fits to the two samples individually are shown in the same colors. A linear fit to all points together is shown in blue. Bottom panel: The correlation between the surface densities of these two quantities.}
\label{hmag}
\end{center}
\end{figure*}

Major mergers are additionally characterized by tidal tails, debris, and other general disruptions of pre-existing galaxy components. The timescales for dissipating these tidal features can be very long (1 Gyr), with longer timescales correlating with larger gas fractions \citep{Lotz2010}. 
Therefore if gas-rich mergers are the primary producer of compact star-forming galaxies at high-redshift, these features should still be present around our compact candidate sample. 
 Since the stellar mass in tidal debris may make up a small fraction of total stellar mass (especially in the case of tidal debris) we estimate here exactly how much stellar mass in extended distributions our stacks account for, given the depth of the stack. This thus provides an upper limit to the amount of residual stellar mass that can be lurking undetected around our galaxy samples.  To estimate this, we take advantage of the relationship between our measured total H-band magnitude and total stellar mass, shown in the top panel of Figure \ref{hmag}. This figure shows that the two are strongly correlated, for all three samples (candidate LBGs, non-candidate LBGs, and compact ETGs) considered here, albeit with large scatter in the case of the total quantities for LBGs (top panel). Nevertheless we can translate this relationship to magnitude per unit area and stellar mass per unit area, to relate surface brightness with surface density of stellar mass. This relationship (for both samples of galaxies together) is shown in the bottom panel of Figure \ref{hmag}. Given the depths of our stacks, which go down to $\sim$ 29 magnitudes per square arcsec, we estimate that the stacks are sensitive to surface densities of at least $10^{6}$ $M_{\odot}$-kpc$^{-2}$. This implies that, if our candidate galaxy samples are compact because of major mergers, the progenitor(s) (i.e. the non-dissipational old stellar component) must contain less than this surface density of old stars in tidal debris or extended stellar halos.

\begin{figure} [!t]
\begin{center}
\includegraphics[scale=0.45]{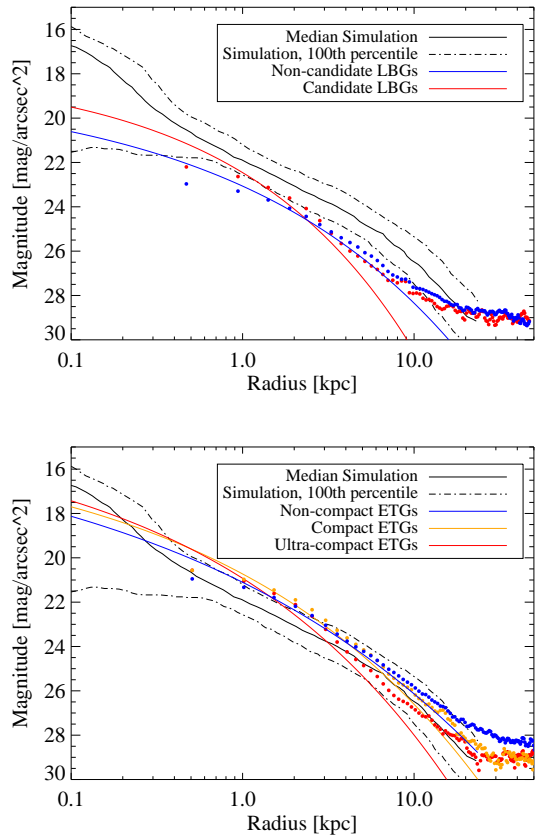}
\caption{Top panel: Intrinsic light profiles from the \citet{Wuyts2010}
  simulated gas-rich merger remnants (median and 100th percentile
  distributions), compared with the observed (points) and intrinsic (i.e. not
  convolved with PSF, lines) sersic profiles of our LBG samples. Bottom panel:
  The same as top panel, with the three samples of ETGs from
  \citet{Cassata2013}}
\label{lastfig}
\end{center}
\end{figure}

\citet{Wuyts2010} present the range of surface brightness profiles of the
extended stellar halos of gas-rich merger remnants, which are more extended
than observed compact ETGs (their Figure 13). We present in Figure
\ref{lastfig} a comparison of these simulated merger remnants, along with the
intrinsic sersic profile (i.e. unconvolved with the PSF) fit to the stacks of
the LBG samples (top panel) as well as the ETG samples (bottom panel).  We
find that in all cases the intrinsic shape of the compact samples (candidate
LBGs, compact and ultra-compact ETGs) do not match the extended distributions
of the simulated merger remnants. While the absolute normalization of the
observed light profiles should depend on galaxy flux, distance, and mass, and
are not necessarily expected to match the simulations in magnitude, the
steepness and shape of the profiles on large scales (i.e. $1<$R$<10$ kpc) do
not agree, independent of this normalization. In other words, the light
profile of the compact galaxies is very often too steep relative to the
extended distributions of the simulated merger remnants, suggesting that if
this extended stellar halo were present in our samples, we would have detected
it. This is in agreement with the comparisons made in \citet{Wuyts2010} to the
observed ETGs of \citet{vanDokkum2008} and \citet{Szomoru2010}, and we show
that similar conclusions can be drawn for our candidate LBG sample. The
extended stellar halo characteristic of simulated gas-rich merger remnants
appears to be absent from our observed light profiles for compact
galaxies. The steepness of the light profiles of these galaxies set important
constraints that any mechanism of compact ETG creation must satisfy, and we
have argued here that this excludes large disky galaxies with shallower
profiles from being compact ETG progenitors.

\subsection{Speculation on a possible scenario for the formation of the compact ETGs and compact galaxies in general}

The above discussion proposes that gas-rich major merging of galaxies with a
sizable stellar component, i.e. observable in the {\it HST} CANDELS images, is
not the mechanism responsible for the formation of the massive, compact ETG
observed at $\langle z\rangle=1.6$. We suggest that a more natural explanation
of our observations may be that the star-formation in these compact galaxies
is being driven primarily by accretion of cold gas, which efficiently forms
stars centrally rather than forming stars in an extended disk. The exact
details of how the gas accretes have been discussed elsewhere by means of
simulations and analytical calculations. It has been suggested that the main
physical mechanism is one where the cold gas dissipates angular momentum in a
compact disk \citep{Danovich2012}, and as more gas accretes the disk develops
VDI \citep{Dekel2009b} that are very effective in driving the gas further down
the bottom of the potential well, giving rise to a very compact
structure. Direct cold mode accretion (CMA) of the cold gas into the compact
structure is also another mechanism \citep{BirnboimDekel2003, Keres2005,
  DekelBirnboim2006} that can give rise to very compact star--forming galaxies
\citep{Johansson2012}. Both the VDI and the CMA predict the formation of very
compact star--forming galaxies, with the VDI-driven wet inflow predicting a
mixture of a perturbed disk and a rotating compact bulge. It is important to
keep in mind that current spectroscopic observations of both compact
star--forming galaxies and passive galaxies \citep{Onodera2012} at $z\sim 2$
do not have sufficient spatial resolution to distinguish between a compact
disk and a spheroid (especially a rotating one), the kinematical signature of
both structures simply being that of broadened emission and absorption
lines. Regardless of the details of how the cold gas is funneled into very
compact regions, the morphology of compact galaxies seem to require a highly
dissipative mechanism for the assembly of their stellar mass as opposed to the
merger of sub--galactic structures with a sizable pre-existing stellar
component.

Star formation in the compact galaxies is then subsequently quenched and they
evolve passively since then. Recent studies have shown that star-formation
can be quenched solely due to feedback from the highly concentrated stellar
distribution. Two examples are stellar winds driven by intense starbursts
\citep{Rupke2005, Tremonti2007, Heckman2011}, and internal ram pressure on
dust grains \citep{Murray2005}. In fact these feedback mechanisms imprint a
maximum possible surface density of star-formation (Eddington limited)
\citep{Murray2005, Thompson2005, Hopkins2010}, 
and evidence of this has been seen in extremely rare compact starbursts at
lower redshifts \citep{DiamondStanic2012}. Other studies have shown that at
high redshift compactness is the most sensitive statistical predictor of
passivity among massive galaxies \citet{Bell2012}, a fact which is in broad
general agreement with the fact that compact and ultra--compact galaxies
dominate the population of passive galaxies at $z>2$, and with the finding
that we have reported here that compact star--forming galaxies appear to have
more evolved bursts compared to non--compact ones.

In conclusion, we speculate here that the high-redshift compact ETGs are the
direct descendants of compact, star-forming galaxies, which themselves are
compact because their star-formation is primarily driven by the accretion of
cold gas to the central regions of the galaxy. Their star-formation is
quenched due to their compactness because of stellar feedback
\citep[e.g.][]{DiamondStanic2012}, halo quenching (in the most massive cases)
\citep{DekelBirnboim2006, Keres2005, BirnboimDekel2003}, or some combination
of both, and evolve passively after. If they undergo merging and or accretion
their compactness is altered and they may end up forming a more diffuse light
profile, and if compact star--forming galaxies do not form anymore, then the
number of compact passive galaxies keeps decreasing with cosmic time (see
Cassata et~al. 2013). This scenario is generally supported by the distribution
of stellar populations we present in Figure \ref{Hprofilesall}, as well as
studies of the main drivers of high-redshift star-formation
\citep{Conselice2013}. With this study we have shown there are
sufficient galaxies to supply the observed abundance of compact ETGs this way,
and that it is not implausible that compact ETGs may be the descendants of
compact star-forming LBGs.

The high-redshift ETG sample of \citet{Cassata2011, Cassata2013} also contain
some fraction of ETGs which are non-compact, (i.e. of size similar to local
ellipticals). At the highest redshifts (z$>$1.5) the fraction is tiny, but the
number density of non-compact ETGs increases dramatically to the present
\citep{Cassata2011, Cassata2013}. Detailed high-resolution studies of local
ellipticals have shown they are best described by multiple morphological
components \citep{Kormendy2009}, even up to three and four sersic components
\citep{Huang2012}, suggesting episodic periods of structural buildup. Other
studies have proposed that the compact ETGs are the cores of local
ellipticals, with stellar mass buildup occurring in an 'inside-out' fashion
\citep{Bezanson2009, Hopkins2009, vanDokkum2010, Huang2012}. What is clear is
that the sizes of ETGs increase dramatically over time, in part because
newly-formed ETGs appear with progressively larger sizes as the universe
evolves \citep{Cassata2011, Cassata2013}. We suggest that these 'non-compact'
z$>$1 ETGs may form from an independent evolutionary track to the compact
ETGs, with non-compact ETGs the result of high-redshift (major) merging
activity.

At z$\gtrsim$1, the number density of non-compact ETGs in the
\citet{Cassata2013} sample steadily increases with time, and has increased
sufficiently to make up half of all M $>$ 10$^{10}$ M$_{\odot}$ ETGs  by
z$\sim$1. Incidentally, the number density of galaxies which are likely to be
gas-rich mergers is similar to that of non-compact ETGs at z$>$1.2 (see Table
\ref{table1}). With a constant merger rate with redshift, and assuming each
merger quenches the star-formation (at least for the next Gyr or two before
another merger rejuvenates star-formation), qualitatively it is plausible that
this steady increase in non-compact ETGs could be explained by a constant
supply of mergers per unit time. In fact, at z$<$1, during the buildup of
these non-compact ETGs, primarily dry mergers are found to account for both
the assembly of the massive end (M$>$10$^{11}M_{\odot}$) of the red sequence
\citep{Vanderwel2009b, Robaina2010}, as well as explain their morphologies as
measured at low-redshift \citep{Vanderwel2009a}. Direct comparison of number
densities to assess whether or not the actual number densities of mergers is
sufficient to explain the increase in non-compact ETGs at z$>$1 will need to
await accurate counts of mergers out to higher redshifts.

\section{Summary}

We have demonstrated the existence of a significant population of compact
LBGs, which have consistent star-formation histories, stellar mass densities,
and co-moving volume densities with high-redshift compact ETGs 
\citep[the sample of][]{Cassata2013}. We find that:
\begin{itemize}
\item These candidate progenitors of compact ETGs show distinct SED properties
  from the non-candidates, consistent with an older burst of SF, i.e. the
  burst appears to show evidence of fading.
\item Stacking from infrared images are consistent with this interpretation,
  and favor an older burst over an increased contribution from dust.

\item The average x-ray properties of the compact and non-compact ones are
  consistent with each other. One interpretation is AGN activity has not
  influenced the selection of the candidates.
\item Structural properties of candidates and compact ETGs differ from
  predictions of gas-rich merger simulations, suggesting this is not the
  dominant mechanism producing compact star-forming galaxies and compact ETGs
  at high-redshift.
\item We suggest the compact ETGs are formed primarily through the quenching
  of compact star-forming galaxies whose in-situ star-formation is driven by
  cold accretion from the IGM, via VDI and CMA. We speculate that merger
  driven evolution may contribute to the non-compact ETG population at
  high-redshift.
\end{itemize}

\acknowledgments

We thank the anonymous referee whose valuable suggestions have improved this paper significantly. MG and CCW wish to thank
Alvio Renzini for many fruitful and inspiring conversations on this work. 
We thank Benjamin Magnelli and Ranga-Ram Chary for graciously providing
24$\mu$m residual maps for our stacking analysis. The
study also benefitted from illustrative discussions with Stacey Alberts,
Alexandra Pope, and Arjen Van der Wel. CCW thanks Joseph Meiring and Dan
Popowich for technical consulting.  This work is based on observations taken
by the CANDELS Multi-Cycle Treasury Program with the NASA/ESA HST.  We
acknowledge support from grant program NSF AST 08-8133, and support for {\it
  HST} Program GO 12060.10-A was provided by NASA through grants from the
Space Telescope Science Institute, which is operated by the Association of
Universities for Research in Astronomy, Inc., under NASA contract NAS5-26555.

\appendix

\section{Possible Quenching Paths for the LBG}
\begin{figure} [!t]
\begin{center}
\includegraphics[scale=0.35]{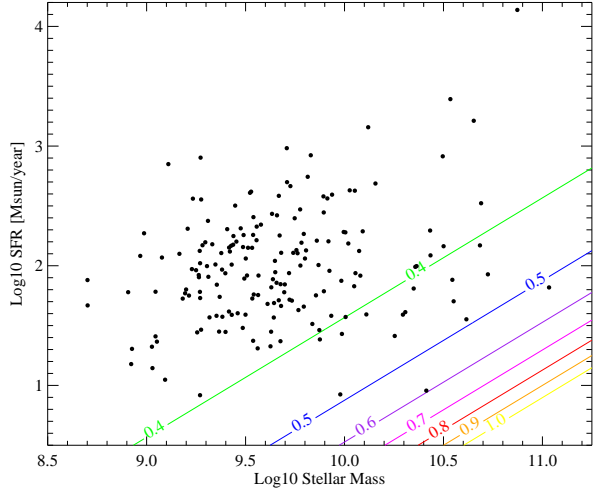}

\caption{This figure shows the limiting value of $\tau$ (in Gyr, contours) for
  exponentially declining SFHs. If the value of $\tau$ is larger than this limiting $\tau$ for a given stellar mass and SFR, then a galaxy at $z\sim3$ will not meet the sSFR criteria of the ETG selection, by $z\sim1.6$. The majority of the parameter space occupied by our galaxies must have $\tau < 0.4$ Gyr.}
\label{maxtau}
\end{center}
\end{figure}

We investigate first the general range of acceptable exponentially declining
SFHs ($\tau$-models) that agree with the red colors and low sSFRs of the
compact ETGs. The low sSFRs of ETGs generally sets a limit to values of the
decline-timescale, $\tau$, and the time since the last starburst activity in
the galaxy. Too recent a burst, or too long a value of the decline timescale,
will leave too high of a sSFR in the galaxy. As the sSFR is a function of both
SFR and also stellar mass, we investigate how the limiting value of $\tau$
changes as a function of both SFR and stellar mass. In Figure \ref{maxtau}, we
show how large $\tau$ can be for a $z \sim 3$ galaxy, as a function of SFR and
stellar mass measured at $z \sim 3$, and still have a measured sSFR $< -2 $
Gyr$^{-1}$ by z$\sim$1.6. This figure shows that, for most LBGs in our sample
(median M* of $10^{9.5} M_{\odot}$, median SFR of $\sim$60 $M_{\odot}$/yr),
the value of $\tau$ must be small ($< 400$Myr). A larger $\tau$ would result
in non-negligible sSFR, and exclusion of the galaxy by the compact ETG
selection.  Not surprisingly, galaxies with lower initial sSFRs (bottom right
corner) can tolerate higher values of $\tau$ and still be considered passive
by z$\sim$1.6.

We estimate a lower-limit to $\tau$ according to the following assumption. For
the star-formation to decline in a galaxy, regardless of the quenching
mechanism, the cold gas must be removed or depleted, and the shortest time in
which this can happen is limited by the sound speed, $c_{s}$, in the ISM of a
star-forming galaxy. Generally, this assumption would imply that each LBG has its own quenching timescale related to the diameter of the galaxy, such that sound crossing-time (and hence quenching time) is larger for larger-sized galaxies. For small galaxies such as the more compact ones we consider here, this quenching time is relatively fast. For a typical sized LBG, diameter D$\sim 2$kpc, we estimate the timescale $t=D/c_{s}=2$kpc$/20$km/s$=100$Myr for cold gas
depletion. In the analysis presented in this paper, we adopt this as our lower limit to $\tau$. However, to test the robustness against this assumption of a fixed minimum timescale for our sample, we note here that repeating the analysis using a sample selected using a different value of $\tau$ for each galaxy that depends on size as outlined above does not significantly change our results. 

With these constraints in hand, our strategy here is to use what we do know
about the SFHs of ETGs, and work backwards to gain some insight into which
LBGs have consistent properties. We acknowledge that any given LBG may or may
not follow an ''ETG-consistent'' SFH between z=3 and z=1.6, but note that any
progenitors of ETGs among the LBGs {\it must} follow an ''ETG-consistent''
one. Therefore, the real ETG progenitors, if any exist among LBGs, will be
contained in a sample selected this way.

We use the lower limit to $\tau$ calculated above, to select our sample, as it
will result in the most conservative sample of plausible progenitors.  By
conservative we mean it results in the smallest number of candidates, with the
smallest increase in their stellar masses over the course of $\sim$2
Gyr. Repeating the analysis with a longer $\tau$ will only cause a net
increase in stellar mass over this time, thus resulting in more massive
galaxies, and additionally adding more candidates. But, as Figure \ref{maxtau}
suggests, alternative SFHs with slightly larger $\tau$'s are consistent with
the observed compact ETG properties, and so the effect of this assumption
should be considered (although the range in allowable values of $\tau$ is
small).

\begin{figure} [!t]
\begin{center}
\includegraphics[scale=0.4]{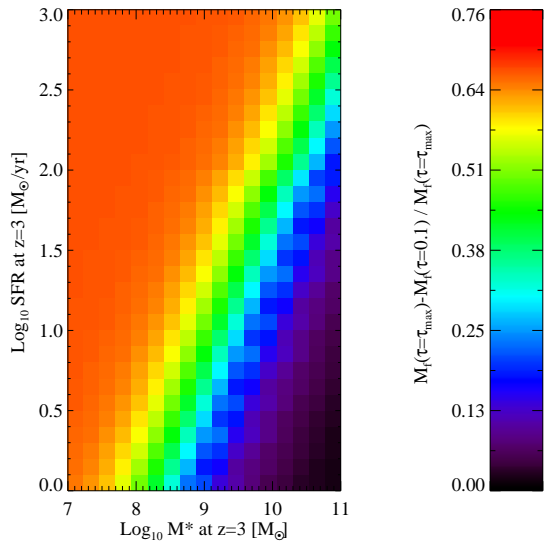}
\caption{Difference in accumulated stellar mass between our assumed value of $\tau$ in section 3 and the maximum $\tau$ allowable in Figure \ref{maxtau}, compared to the maximum amount. The difference between the two is somewhat comparable to the uncertainties in stellar mass measurements in \citet{JLee2009}. }
\label{deltam}
\end{center}
\end{figure}

Phrased in another way, this range in $\tau$ introduces a scatter in the {\it
  actual} amount of stellar mass that LBGs will form (compared to that which
we {\it assume}), between $z\sim3$ and $z\sim1.6$. In Figure \ref{deltam}, we
assess exactly how much this scatter in stellar mass buildup is between our
upper and lower limits in $\tau$. To quantify the scatter, we calculate
$\frac{\Delta M(\tau)}{M}=\frac{M(\tau_{max}) - M(100Myr)}{M(\tau_{max})}$ for
a range of initial values of SFR and $M_{*}$ measured at $z\sim3$, analogous
to Figure \ref{maxtau}. For each region of the figure we use as the maximum
value of $\tau$ the value derived in Figure \ref{maxtau}.  We find that the
assumption of different $\tau$ values within the limits have less than a
factor of two effect on the change in mass, compared to final mass, of the
LBGs. We can compare these values to the actual uncertainty in the estimate of
stellar mass in LBGs, for example, according to the simulations of stellar
mass estimates for LBGs in the analysis of \citet{JLee2009}. The estimated
error in stellar mass of LBGs is magnitude dependent, but may vary up to a
factor of 2 for U--band dropouts. The scatter we find in final mass, depending
on assumption of the SFH, is less than that scatter associated with the
initial LBG mass estimate. Therefore, the differences in extrapolated mass
buildup from varying the assumed value of $\tau$ are not the dominant source
of uncertainty in weather or not an LBG is included in the candidate sample. We additionally note that we do not make an attempt to incorporate mass-loss from stars that is returned to the ISM during the cycle of star-formation.

\end{document}